\documentclass{amsart}
\usepackage{amssymb}
\usepackage{amsfonts}

\newtheorem{theorem}{Theorem}
\theoremstyle{plain}

\newtheorem{lemma}{Lemma}

\numberwithin{equation}{section}
\input{tcilatex}

\begin{document}
\title[Quantum Continual Measurement of CCR]{Quantum Continual Measurements
and a Posteriori Collapse on CCR}
\author{V. P. Belavkin}
\address{Mathematics Department, University of Nottingham,UK}
\email{vpb@maths.nott.ac.uk}
\thanks{On leave of absence from M.I.E.M., B.Vusovski Street 3/12 Moscow
109028, USSR}
\date{Received: October 2, 1990}
\dedicatory{}
\thanks{This paper is published in: \textit{Communications Mathematical
Physics} \textbf{146}, pp 611--635 (1992).}

\begin{abstract}
A quantum theory for the Markovian dynamics of an open system under the
unsharp observation which is continuous in time, is developed within the CCR
stochastic approach. A stochastic classical equation for the posterior
evolution of quantum continuously observed system is derived and the
spontaneous collapse (stochastically continuous reduction of the wave
packet) is described. The quantum Langevin evolution equation is solved for
the general linear case of a quasi--free Hamiltonian in the initial CCR
algebra with a fixed output observable field, and the posterior Kalman
dynamics coresponding to an initial Gaussian state is found. It is shown for
an example of the posterior dynamics of quantum unstable open system that
any mixed state under a complete nondemolition measurement collapses
exponentially to a pure Gaussian one.
\end{abstract}

\maketitle

\section{Introduction}

The time evolution of quantum system under an observation which is
continuous in time cannot be described by any Schr\"{o}dinger equation due
to the stochastic irreversible nature of von Neumann reduction of the wave
packet at any instant of measurement. An adequate model of the quantum
unitary evolution giving a continuous collapse by a conditioning with
respect to the measurements can be obtained in the framework of quantum
stochastic (QS) calculus \cite{bib:12}, firstly introduced for output
nondemolition processes in \cite{bib:ref2,bib:b1} and recently developed in
a quite general form in \cite{bib:bel1,bib:2,bib:b7}. A stochastic wave
equation for an observed quantum system derived in \cite{bib:b7} by using
the quantum filtering method \cite{bib:2}, provides an explanation of pure
quantum relaxation of an atom under a complete observation \cite{bib:bel2}
(Zeno paradox) and a Watch--Dog effect \cite{bib:bel3} for the reduced wave
function of a quantum particle under the continuous observation.

In this paper we develop a regorous quantum stochastic theory of unsharp
nondemolition measurements of continual families of arbitrary noncommuting
observables $R_{t,\mathbf{x}}$ given sequentially in the real space-time $(t,%
\mathbf{x})\in \mathbb{R}^{1+d}$. In the case $d=0$ this defines the
standard unitary dilation of an instrumental process for the quantum
measurements, which are continuous in time, considered within an operational
approach by Barchielli and Lupieri \cite{bib:bel1}. We give the direct proof
of stochastic evolution equation for the posterior states of a general
quantum system under a continuous indirect measurement of a noncommutative
field-process $\boldsymbol{R}_{t}=\{R_{t,\mathbf{x}}\mid \mathbf{x}\in 
\mathbb{R}^{d}\}$. The observed process $Y(t)$ is supposed to be
nondemolition in the sense \cite{bib:2} -- \cite{bib:bel2} of the
commutativity $[Y(r),\;X(t)]=0$ of the past observables $Y^{t}=\{Y(r)|r\leq
t\}$ with the Heisenberg operators $X(t)$ of the system for every $t$ and
self-nondemolition (commutative) $[Y(r),\;Y(t)]=0$ for all $r,\;t$. In that
case a posteriori state can be found \cite{bib:bel3} for any initial prior
state by the Takesaki conditional expectations $\epsilon \{X(t)|Y^{t}\}$ on $%
\{Y(r)\mid r\leq t\}^{\prime }$ restricted tothe future von Neumann algebras 
$\mathcal{L}_{t}=\{X(s)|s\geq t\}^{\prime \prime }$. We shall show that it
is possible to represent the open quantum system under observation within a
class of quantum stochastic evolutions in such a way that the observed
commutative process $Y(t)$ for the sequential unsharp measurements of a
noncommutative process $R_{t}$ is described as the sum of noncommutative
Heisenberg operators $R(t)=U^{\ast }(t)R_{t}U(t)$ of the subsystem under the
measurement and a classical (commutative) while noise (error) $e(t)$. The
unitary evolution $U(t)$ of such systems perturbed by a singular interaction
with a meter is described in a 'Bose reservoir' by a quantum stochastic Schr%
\"{o}dinger equation \cite{bib:12}, driven by a white noise (force) $f(t)$.
Note that the force $f(t)$ responsible for the perturbation of the system
due to the measurements, may appear in the quantum Langevin equation as well
as the classical (commutative) white noise \cite{bib:2}--\cite{bib:bel2}.
But the pair $(e,\,f)$ cannot be described within the classical theory of
generalized processes any more because the error $e(t)$ does not commute
with $f(t)$ given the nondemolition condition for $R(t)$ and $Y(t)=R(t)+e(t)$%
.

It is interesting to note that stochastic equations of the particular
diffusive type of~(\ref{eq:ccr2.3}) and~(\ref{eq:ccr2.9}), in their
normalized nonlinear version \cite{bib:2,bib:b7,bib:bel2}, have appeared in
the physical literature also in connection with phenomenological dynamical
theories of quantum reduction and spontaneous collapse \cite{bib:bel6} --- 
\nocite{bib:bel7,bib:bel8,bib:bel9}\cite{bib:bel10}. The idea is that the
wave-function reduction associated to a continual measurement is some kind
of diffusion process and some particular equations of this type are
postulated. Our approach shows that this diffusion postulate as well as the
continual counting reduction \cite{bib:ref33} can be derived in the natural
general form from the unitary stochastic evolution of a big quantum system
by the conditioning with respect to a chosen nondemolition process under the
continual measurement. The unsharp self-nondemolition measurements and the
objectification problem are discussed now intensively in the physical
literature \cite{bib:bel12, bib:bel13} within the Davies-Lewis-Ludwigs
operational approach, but real progress in clarifying the connection between
the operational theory of continual measurements \cite{bib:bel1} and the
spontaneous reduction theories \cite{bib:bel6} -- \cite{bib:bel8} can be
done only by using the quantum stochastic and nonlinear filtering methods 
\cite{bib:2} -- \cite{bib:bel2},\cite{bib:ref33} which are considered in
this paper.

\section{A quantum stochastic model with continual unsharp measurements}

Let us consider the dynamical problem of a sequential observation in
continuous time $t\geq 0$ of a measurable family $\boldsymbol{L}_{t}=\{L_{t,%
\mathbf{x}}|\mathbf{x}\in \Lambda \}$ of operators $L_{x}=L_{t,\mathbf{x}%
},\;x=(t,\mathbf{x})$ in a Hilbert space $\mathcal{H}$, where $\Lambda $ is
a Borel space with a $\sigma $-algebra $\mathcal{A}$. We do not suppose that
the operators $L_{x}$ are pairwise commutative or even self--adjoint or
normal. But we at first assume that they are bounded, $L_{x}\in \mathcal{L}(%
\mathcal{H})$, almost everywhere on the space $\mathbb{R}_{+}\times \Lambda $
with respect to the product $\lambda (\mathrm{d}x)=\mathrm{d}t\lambda (%
\mathrm{d}\mathbf{x})$ of a positive measure $\lambda (\mathrm{d}\mathbf{x})$
on the Borel space $\Lambda $ and the standard Lebesque measure $\mathrm{d}t$
on $\mathbb{R}_{+}$. Here $\mathcal{L(H)}$ denotes the space of continuous
(bounded) operators in $\mathcal{H}$.

One can consider for example the problem of the (indirect) measurement of
spin momenta $L_{t,\mathbf{x}}=L_{\mathbf{x}}$, described in the Schr\"{o}%
dinger picture by the operators in $\mathcal{H}=\mathbb{C}^{2}$ of spin
projections $L_{\mathbf{x}}=\frac{1}{2}\ M(\mathrm{d}\mathbf{x})/\mathrm{d}%
\mathbf{x}+L_{\mathbf{x}}^{\ast }$, where $M(\Delta )=\int_{\Delta }R_{%
\mathbf{x}}\mathrm{d}\mathbf{x}$ is an operator--valued measure $M(\Delta
)\in \mathcal{L}(\mathbb{C}^{2})$ of the momentum in a solid angle $\Delta
\subset \Lambda $ with $R_{\mathbf{x}}=L_{\mathbf{x}}+L_{\mathbf{x}}^{\ast }$
having the eigen--values $\pm 1$, and $\lambda (\mathrm{d}\mathbf{x})=%
\mathrm{d}\mathbf{x}$ is the standard solid angle measure on the sphere $%
\Lambda =\{\mathbf{x}\in \mathbb{R}^{3}||\mathbf{x}|=1\}$, normalized to $%
4\pi $.

Due to the absence of a joint spectral resolution for the noncommutative
family $\{L_{\mathbf{x}}|\mathbf{x}\in \Lambda \}$, there is no possibility
of measuring the corresponding physical quantities in the usual (direct)
sense. Moreover there is no way within orthodox quantum mechanics and
measurement theory to describe an observation which is continuous in time
even for a single self--adjoint operator $L_{\mathbf{x}}=L$ with a simple
spectrum or to predict the dynamics of the quantum system under such an
observation due to the absence of nontrivial mathematical models for
noninstantaneous measurements.

We shall show that these difficulties can be removed within the quantum
theory of open systems and indirect measurements, based on the quantum
stochastic approach \cite{bib:12,bib:b1}. The basic idea is that the quantum
system under an observation must be described as the subsystem of a big
system, including a Boson field $\boldsymbol{A}$ as a model of an
observation channel coupled to the system by a singular interaction. The
measurement information about the physical quantities $L_{x}$ under such
coupling can be continuously extracted in a nondemolition way from the
continually-sequential unsharp observation of the output field $\boldsymbol{B%
}=U_{\infty }\boldsymbol{A}U_{\infty }^{\ast }$ given by the direct
measurements of the compartible complex observables $\boldsymbol{Z}=${$%
\boldsymbol{B}$}${+}${$\boldsymbol{A}$}$_{\circ }^{\ast }$.

Let $A(\mathrm{d}x)$ be the Bose-field annihilation measure on $\Gamma _{1}=%
\mathbb{R}_{+}\times \Lambda $, satisfying the canonical commutation
relations (CCR) 
\begin{equation}
\lbrack A(\Delta ^{\prime }),\,A^{\ast }(\Delta )]=\lambda (\Delta \cap
\Delta ^{\prime }),\;\;\forall \Delta ,\,\Delta ^{\prime }\in \mathcal{A}%
(\Gamma _{1})  \label{eq:ccr1.1}
\end{equation}%
in the Fock space $\mathcal{F}$ over the Hilbert space $L^{2}(\Gamma _{1})$
of square integrable functions of $x\in \Gamma _{1}$. One can realize \cite%
{bib:bel5} $\mathcal{F}$ as the space $L^{2}(\Gamma )=\oplus _{n=0}^{\infty
}L^{2}(\Gamma _{n})$ of functions $f$, square integrable in the sense that 
\begin{equation}
\int |f(\chi )|^{2}\lambda (\mathrm{d}\chi )=\sum_{n=0}^{\infty }\int \ldots
\int_{0\leq t_{n}<\ldots <t_{n}<\infty }|f(x_{1},\ldots
,x_{n})|^{2}\prod_{i=1}^{n}\lambda (\mathrm{d}x_{i})<\infty ,
\label{eq:ccr1.2}
\end{equation}%
of chains $\chi =(x_{1},\ldots ,x_{n}),\;x_{i}=(t_{i},\mathbf{x}%
_{i}),\;t_{1}<\ldots <t_{n}$ of all finite lengths $|\chi |=n=0,\,1,\ldots $
with respect to the natural measure $\lambda (\mathrm{d}\chi )=\prod_{x\in
\chi }\lambda (\mathrm{d}x)$. We identify the chains $\chi \in \Gamma $ as
subsets $\{x_{1},\ldots ,x_{n}\}\subset \Gamma _{1},\,t_{1}<\ldots <t_{n}$
and the time ordered elements $(x_{1},\ldots ,x_{n})\in \Gamma _{1}^{n}$ of
the $n$-cube $\Gamma _{1}^{n}$, so that $\Gamma =\bigcup_{n=o}^{\infty
}\Gamma _{n}$ is considered as the direct union of the sets $\Gamma
_{n}=\{(x_{1},\ldots ,x_{n})|t_{1}<\ldots <t_{n}\}$. Then the annihilation
operator $A(\Delta )$ of the Boson quanta in a measurable region $\Delta \in
\Gamma _{1}$ is 
\begin{equation}
\lbrack A(\Delta )\,f](\chi )=\int_{\Delta }f(\chi \sqcup x)\lambda (\mathrm{%
d}x),  \label{eq:ccr1.3}
\end{equation}%
where $\chi \sqcup x$ is defined as the chain $(x_{1},\ldots
,x_{i},x,x_{i+1},\ldots ,x_{n})$ of length $n+1$ for almost all $x=(t,%
\mathbf{x})$, namely if $t\notin \{t_{1},\ldots ,t_{n}\}$.

One can easily find that the operator~(\ref{eq:ccr1.3}) is adjoint to the
creation operator $A^{\ast }(\Delta )$ of the quanta in $\Delta \in \mathcal{%
A}(\Gamma _{1})$ 
\begin{equation}
(A^{\ast }(\Delta )f)(\chi )=\sum_{x\in \chi (\Delta )}f(\chi \backslash
x),\;\;\chi (\Delta )=\chi \cap \Delta ,  \label{eq:ccr1.4}
\end{equation}%
with respect to the scalar product~(\ref{eq:ccr1.2}) and satisfies the CCR~(%
\ref{eq:ccr1.1}), where $\chi \backslash x=(x_{1},\ldots
,x_{i-1},x_{i+1},\ldots ,x_{n})$ is the complement of the elementary chain $%
x\in \Gamma _{1}$ in the chain $\chi \in \Gamma _{n}$ with $x_{i}=x\in \chi $%
. In the following we shall regard the operators $A(\Delta )$, $A^{\ast
}(\Delta )$ acting as~(\ref{eq:ccr1.3}),~(\ref{eq:ccr1.4}) in the Hilbert
space $\mathcal{H}\otimes \mathcal{F}$ of square integrable
vector--functions $h:\Gamma \rightarrow \mathcal{H}$ with the invariant
domain $\mathcal{D}=\bigcup_{\xi >1}\mathcal{D}(\xi )$, where 
\begin{equation*}
\mathcal{D}(\xi )=\{h\in \mathcal{H}\otimes \mathcal{F}|\;\int \xi ^{|\chi
|}\Vert h(\chi )\Vert ^{2}\lambda (\mathrm{d}\chi )<\infty \}.
\end{equation*}

Let us consider the quantum stochastic evolution $U_{t}$, $t\in \mathbb{R}%
_{+}$ in $\mathcal{H}\otimes \mathcal{F}$, given by the
Hudson--Parthasarathy operator equation \cite{bib:12,bib:bel5} $\mathrm{d}%
U+KU\mathrm{d}t=(\boldsymbol{L}\mathrm{d}\boldsymbol{A}^{\ast }-\boldsymbol{L%
}^{\ast }\mathrm{d}\boldsymbol{A})U$ for $U(t)=U_{t}^{\ast }$ having in our
(nonstationary) case the form 
\begin{equation}
\mathrm{d}U_{t}^{\ast }+K_{t}U_{t}^{\ast }\mathrm{d}t=\int_{\Lambda }(%
\mathrm{d}A^{\ast }(t,\mathrm{d}\mathbf{x})L_{x}-L_{x}^{\ast }\mathrm{d}A(t,%
\mathrm{d}\mathbf{x}))U_{t}^{\ast },\;\;U_{0}^{\ast }=I,  \label{eq:ccr1.5}
\end{equation}%
where $K_{t}=iH_{t}+\frac{1}{2}\ \int L_{t,\mathbf{x}}^{\ast }L_{t,\mathbf{x}%
}\lambda (\mathrm{d}\mathbf{x})$, the integral is taken over $\mathbf{x}\in
\Lambda $, $A(t,\mathrm{E})=A([0,t)\times \mathrm{E})$, and $\mathrm{d}A(t,%
\mathrm{E})=A(t+\mathrm{d}t,\mathrm{E})-A(t,\mathrm{E})$ is the forward
differential of the process $A(t,\mathrm{E})$ for fixed $\mathrm{E}\in 
\mathcal{A}$. The necessary condition for the unitarity $U_{t}^{\ast
}=U_{t}^{-1}$ of the family $U_{t},\;t>0$ satisfying the quantum stochastic
differential equation~(\ref{eq:ccr1.5}) is \cite{bib:12} the
self--adjointness of the operators $H_{t}$ (Hamiltonian) in $\mathcal{H}$
and that the integrals $\int_{\Lambda }L_{t,\mathbf{x}}^{\ast }L_{t,\mathbf{x%
}}\lambda (\mathrm{d}\mathbf{x})$ exist and equal $K_{t}+K_{t}^{\ast }$.

The solution of equation~(\ref{eq:ccr1.5}) can be described \cite{bib:bel5}
explicitly in terms of the quantum stochastic multiple integral in Fock
scale provided the conditions 
\begin{equation}
\int_{t<s}\Vert H_{t}\Vert \mathrm{d}t<\infty ,\;\int_{t<s}\int_{\Lambda
}\Vert L_{t,\mathbf{x}}\Vert ^{2}\mathrm{d}t\lambda (\mathrm{d}\mathbf{x}%
)<\infty ,\;\;\forall s\in \mathbb{R}_{+}  \label{eq:ccr1.6}
\end{equation}%
hold which are sufficient for the existence and uniqueness of the unitary
solution $U_{t}$ of equation~(\ref{eq:ccr1.5}) with $H_{t}=H_{t}^{\ast }$.

Let us define the output observed process $\boldsymbol{Y}(t)$ of unsharp
measurements of the continual family $\{L_{x}+L_{x}^{\ast }\mid x\in \Gamma
_{1}\}$ as the time dependent selfadjoint operator--valued measure $Y(t,%
\mathrm{E}),\;\mathrm{E}\in \mathcal{B}$ on some $\sigma $-semi-ring $%
\mathcal{B}\subseteq \mathcal{A}$ with $\lambda (\mathrm{E})<\infty $ given
by the quantum stochastic (forward) differential 
\begin{equation}
\mathrm{d}Y(t,\mathrm{E})=M(t,\mathrm{E})\mathrm{d}t+\mathrm{d}Q(t,\mathrm{E}%
),\;\;Y(0,\mathrm{E})=0,  \label{eq:ccr1.7}
\end{equation}%
where $M(t,\mathrm{E})=\int_{\mathrm{E}}U_{t}(L_{t,\mathbf{x}}+L_{t,\mathbf{x%
}}^{\ast })U_{t}^{\ast }\lambda (\mathrm{d}\mathbf{x})$, $Q(t,\mathrm{E}%
)=A(t,\mathrm{E})+A^{\ast }(t,\mathrm{E})$. In the case of the initial
vacuum state $|0\rangle \in \mathcal{F}$ of the Bose field and $\mathcal{B}$
generating $\mathcal{A}$, the generalized processes $\dot{Y}_{\mathbf{x}}(t)=%
\mathrm{d}Y(t,\mathrm{d}\mathbf{x})/\mathrm{d}t\lambda (\mathrm{d}\mathbf{x}%
) $ can be regarded as a complete indirect observation of noncommuting
operators $R_{x}=L_{x}+L_{x}^{\ast },\;x\in \Gamma _{1}$ given by the
instantaneous sequential~measurements of the commuting operators $Y(\mathrm{d%
}x)=\mathrm{d}Y(t,\mathrm{d}\mathbf{x})$. Indeed the differentials $\mathrm{d%
}Q(t,\mathrm{E})$ for all measurable $\mathrm{E}\in \Lambda $ in that case
are statisticaly equivalent to Wiener increments with zero mean values $%
\langle 0|\mathrm{d}Q(t,\mathrm{E})|0\rangle =0$ and minimal covariances $%
\langle 0|\mathrm{d}Q(t,\mathrm{E}^{\prime })\mathrm{d}Q(t,\mathrm{E}%
)|0\rangle =\mathrm{d}t\lambda (\mathrm{E}\cap \mathrm{E}^{\prime })$
compatible with the CCR~(\ref{eq:ccr1.1}). They are independent of the
operators $M(t,\mathrm{E})=\int_{\mathrm{E}}R_{\mathbf{x}}(t)\lambda (%
\mathrm{d}\mathbf{x})$, defined at the infinitesimal volume $\mathrm{E}=%
\mathrm{d}\mathbf{x}$ by the Heisenberg operators $R_{\mathbf{x}%
}(t)=U_{t}R_{x}U_{t}^{\ast }$ as $M(t,\mathrm{d}\mathbf{x})=R_{\mathbf{x}%
}(t)\lambda (\mathrm{d}\mathbf{x})$. Hence the differences between the
increments $\mathrm{d}Y(t,\mathrm{d}\mathbf{x})=Y(t+\mathrm{d}t,\mathrm{d}%
\mathbf{x})-Y(t,\mathrm{d}\mathbf{x})$ of the form~(\ref{eq:ccr1.7}) and the
operators $R_{\mathbf{x}}(t)\mathrm{d}t\lambda (\mathrm{d}\mathbf{x})$ are
just independent Gaussian variables $\mathrm{d}Q(t,\mathrm{d}\mathbf{x})$,
defining the minimal random error of the measurement of the noncommutative
family $\boldsymbol{R}_{t}=\{R_{t,\mathbf{x}}|\mathbf{x}\in \Lambda \}$ in
the continuous time $t\in \mathbb{R}_{+}$ as white noise $\dot{\boldsymbol{Q}%
}(t)=\{\dot{Q}_{\mathbf{x}}(t)|\mathbf{x}\in \Lambda \}$. One can consider $%
Y(t,\mathrm{E}),\;\mathrm{E}\in \mathcal{B}$ as a coarse--graining $%
Y_{i}(t)=Y(t,\mathrm{E}_{i})$ of the family $Y(t,\mathrm{E}),\;\mathrm{E}\in 
\mathcal{A}$, corresponding to a $\sigma $-partition $\mathcal{B}=\{\mathrm{E%
}_{i}\in \mathcal{A}\mid i\in I\}$ of a measurable subset $\mathrm{M}=\sum 
\mathrm{E}_{i}\subseteq \Lambda $.

The following theorem shows that the QS equation (\ref{eq:ccr1.5}) up to the
Hamiltonian $H_{t}$ corresponds to the unique Evans--Hudson diffusion $%
j(t,X)=U_{t}XU_{t}^{\ast }$, satisfying the nondemolition principle 
\begin{equation*}
\lbrack X(s),B(t,\mathrm{E})],\;\;\forall t\leq s\in \mathbb{R}_{+},\;X\in 
\mathcal{L},\;\mathrm{E}\in \mathcal{A}
\end{equation*}%
for all $X(t)=j(t,X)$ over a von-Neumann initial subalgebra $\mathcal{L}%
\subseteq \mathcal{L}(\mathcal{H})$ respectively to the given output field 
\begin{equation*}
B(t,\mathrm{E})=\int_{0}^{t}\int_{\mathrm{E}}L_{x}\left( r\right) \lambda (%
\mathrm{d}\mathbf{x})\mathrm{d}r+A(t,\mathrm{E}),\;\;\mathrm{E}\in \mathcal{A%
}.
\end{equation*}

\begin{theorem}
Let $j(t):\mathcal{L}\rightarrow \mathcal{L}(\mathcal{H}\otimes \mathcal{F})$%
, $t\in \mathbb{R}_{+}$ be a quantum diffusion over a unital $\ast $-algebra 
$\mathcal{L}\subseteq \mathcal{L}(\mathcal{H})$ having the QS-differential 
\begin{equation*}
\mathrm{d}j(t,X)=\gamma (t,X)\mathrm{d}t+\boldsymbol{\Delta }(t,X)\mathrm{d}%
\boldsymbol{A}^{\ast }(t)+\boldsymbol{\Delta }^{\ast }(t,X)\mathrm{d}%
\boldsymbol{A}(t),
\end{equation*}%
where $\mathrm{d}\boldsymbol{A}^{\ast }\boldsymbol{\Delta }=\int_{\Lambda }%
\mathrm{d}A^{\ast }(\mathrm{d}\mathbf{x})\delta _{x}$, $\boldsymbol{\Delta }%
^{\ast }\mathrm{d}\boldsymbol{A}=\int_{\Lambda }\delta _{x}\mathrm{d}A(%
\mathrm{d}\mathbf{x})$, and%
\begin{equation*}
\delta _{x}(t,X^{\ast })=\delta _{x}^{\ast }(t,X)^{\ast },\gamma (t,X^{\ast
})=\gamma (t,X)^{\ast }
\end{equation*}%
are the linear structural maps $\mathcal{L}\rightarrow \mathcal{L}(\mathcal{H%
}\otimes \mathcal{F})$, necessary satisfying the conditions 
\begin{eqnarray*}
(i) &&\delta _{x}(t,X^{\ast }X)=j(t,X^{\ast })\delta _{x}(t,X)+\delta
_{x}^{\ast }(t,X^{\ast })j(t,X), \\
(ii) &&\delta _{x}^{\ast }(t,X^{\ast }X)=j(t,X^{\ast })\delta _{x}^{\ast
}(t,X)+\delta _{x}(t,X^{\ast })j(t,X), \\
(iii) &&\gamma (t,X^{\ast }X)=\boldsymbol{\Delta }(t,X)^{\ast }\boldsymbol{%
\Delta }(t,X)-j(t,X)^{\ast }\gamma (t,X)-\gamma (t,X)^{\ast }j(t,X),
\end{eqnarray*}%
with $\boldsymbol{\Delta }(X)^{\ast }\boldsymbol{\Delta }(X)=\int_{\Lambda
}\delta _{x}(X)^{\ast }\delta _{x}(X)\lambda (\mathrm{d}\mathbf{x})$, $%
\delta _{x}(t,I)=0=\delta _{x}^{\ast }(t,I)$, $\gamma (t,I)=0$. The family $%
\{X(t)=j(t,X)\mid X\in \mathcal{L}\}$ satisfies the nondemolition condition 
\begin{equation}
\lbrack X(s),Y(t,\mathrm{E})]=0,\;\;\forall s\leq t  \label{eq:ccr1.8}
\end{equation}%
with respect to the output fields $Y(t)\in \{B(t,\mathrm{E}),B^{\ast }(t,%
\mathrm{E})\}$, 
\begin{eqnarray*}
\mathrm{d}B(t,\mathrm{E}) &=&\int_{\mathrm{E}}j(t,L_{x})\lambda (\mathrm{d}%
\mathbf{x})\mathrm{d}t+\mathrm{d}A(t,\mathrm{E}),\;\;\mathrm{E}\in \mathcal{A%
}, \\
\mathrm{d}B^{\ast }(t,\mathrm{E}) &=&\int_{\mathrm{E}}j(t,L_{x}^{\ast
})\lambda (\mathrm{d}\mathbf{x})\mathrm{d}t+\mathrm{d}A^{\ast }(t,\mathrm{E}%
),\;\;t\in \mathbb{R}_{+}
\end{eqnarray*}%
if and only if $\boldsymbol{\Delta }$, $\boldsymbol{\Delta }^{\ast }$ are
the inner differentiations: 
\begin{eqnarray*}
\delta _{x}(t,X) &=&j(t,[X,L_{t,\mathbf{x}}]),\;\;\delta _{x}^{\ast
}(t,X)=j(t,[L_{t,\mathbf{x}}^{\ast },X]), \\
\gamma (t,X) &=&\beta (t,X)+\frac{1}{2}\ \int_{\Lambda }j(t,L_{x}^{\ast
}[X,L_{x}]+[L_{x}^{\ast },X]L_{x})\lambda (\mathrm{d}\mathbf{x}),
\end{eqnarray*}%
where $\beta (t,X^{\ast })=\beta (t,X)^{\ast }$ is a $j(t)$-differentiation $%
\mathcal{L}\rightarrow \mathcal{L}(\mathcal{H}\otimes \mathcal{F})$: 
\begin{equation*}
\beta (t,X^{\ast }X)=j(t,X^{\ast })\beta (t,X)+\beta (t,X^{\ast
})j(t,X),\;\;\beta (t,I)=0.
\end{equation*}%
In the inner case $\beta (t,X)=j(t,i[H_{t},X])$ these conditions together
with (\ref{eq:ccr1.6}) uniquelly define the quantum Markov spatial flow $%
j(t,X)=U_{t}XU_{t}^{\ast }$ given by the Hudson--Parthasarathy equatuion (%
\ref{eq:ccr1.5}). Moreover, the output fields $\boldsymbol{B}(t)$, $%
\boldsymbol{B}^{\ast }(t)$ and, hence, the nondemolition process $%
\boldsymbol{Y}(t)=\boldsymbol{B}(t)+\boldsymbol{B}^{\ast }(t)$ are locally
unitary equivalent to the input fields $\boldsymbol{A}(t)$, $\boldsymbol{A}%
^{\ast }(t)$ and to the commutative process $\boldsymbol{Q}(t)=\{Q(t,\mathrm{%
E})\mid \mathrm{E}\in \mathcal{B}\}$: $\boldsymbol{B}(t)=U_{\infty }%
\boldsymbol{A}(t)U_{\infty }^{\ast }$, $\boldsymbol{B}^{\ast }(t)=U_{\infty }%
\boldsymbol{A}^{\ast }(t)U_{\infty }^{\ast }$ in the sense 
\begin{equation}
Y(t,\mathrm{E})=U_{s}Q(t,\mathrm{E})U_{s}^{\ast },\;\;\;\forall s\geq t.
\label{eq:ccr1.9}
\end{equation}%
In particular, $Y(t,\mathrm{E})=U_{t}Y_{t}(\mathrm{E})U_{t}^{\ast }$ for all 
$\mathrm{E}\in \mathcal{B}$, where $Y_{t}(\mathrm{E})=Q([0,t)\times \mathrm{E%
})=Q(t,\mathrm{E})$, $Q(\mathrm{d}x)=A(\mathrm{d}x)+A^{\ast }(\mathrm{d}%
x),\;x\in \Gamma _{1}$. \label{th:a1}
\end{theorem}

\textbf{Proof.} The increments $\mathrm{d}X(t)=X(t+\mathrm{d}t)-X(t)$ of the
linear $\ast $-maps $j(t):X\mapsto X(t)$, $j(t,X)^{\ast }=j(t,X)^{\ast }$
uniquelly define the linear $\ast $-map $\gamma (t):\mathcal{L}\rightarrow 
\mathcal{L}(\mathcal{H}\otimes \mathcal{F})$ and the adjoint maps $%
\boldsymbol{\Delta }(t)$, $\boldsymbol{\Delta }^{\ast }(t)$ due to the
linear independence of the differentials $\mathrm{d}t$ and $\mathrm{d}%
\boldsymbol{A}^{\ast }(t)$, $\mathrm{d}\boldsymbol{A}(t)$. An application of
the QS Ito formula \cite{bib:12} to the conditions $j(t,I)=I$, $j(t,X^{\ast
}X)=j(t,X)^{\ast }j(t,X)$ gives $\gamma (t,I)=0$, $\boldsymbol{\Delta }%
(t,I)=0=\boldsymbol{\Delta }^{\ast }(t,I)$, and 
\begin{eqnarray*}
\mathrm{d}(X(t)^{\ast }X(t)) &=&\mathrm{d}X(t)^{\ast }\mathrm{d}X(t)+\mathrm{%
d}X(t)^{\ast }X(t)+X(t)^{\ast }\mathrm{d}X(t) \\
&=&[\boldsymbol{\Delta }(t,X)^{\ast }\boldsymbol{\Delta }(t,X)-\gamma
(t,X)^{\ast }j(t,X)-j(t,X)^{\ast }\gamma (t,X)]\mathrm{d}t \\
&&+(\boldsymbol{\Delta }^{\ast }(X)j(X)+j(X)^{\ast }\boldsymbol{\Delta }%
^{\ast }(X))\mathrm{d}\boldsymbol{A} \\
&&+(\boldsymbol{\Delta }(X^{\ast })j(X)+j(X^{\ast })\boldsymbol{\Delta }(X))%
\mathrm{d}\boldsymbol{A}^{\ast }.
\end{eqnarray*}%
Comparing this with the QS differential 
\begin{equation*}
\mathrm{d}j(t,X^{\ast }X)=\boldsymbol{\Delta }^{\ast }(X^{\ast }X)\mathrm{d}%
\boldsymbol{A}+\boldsymbol{\Delta }(X^{\ast }X)\mathrm{d}\boldsymbol{A}%
^{\ast }-\gamma (X^{\ast }X)\mathrm{d}t,
\end{equation*}%
we obtain the conditions (i), (ii), (iii), found in \cite{bib:lis30} for the
Markovian case.

If $Y(t)$ is a nondemolition process respectively to $X(t)$, then 
\begin{equation*}
\lbrack \mathrm{d}X(t),Y(s)]=[X(t+\mathrm{d}t),Y(s)]-[X(t),Y(s)]=0,\;\;%
\forall t\geq s
\end{equation*}%
for $t\geq s$ and hence 
\begin{equation*}
\lbrack \gamma (t,X),Y(s)]=[\delta _{x}(t,X),Y(s)]=[\delta _{x}^{\ast
}(t,X),Y(s)]=0,\;\;\forall t\geq s
\end{equation*}%
due to the commutativity of $\mathrm{d}t$, $\mathrm{d}A^{\ast }(t,\mathrm{E}%
) $, $\mathrm{d}A(t,\mathrm{E})$ with $Y(s)$, $s\leq t$. Applying the QS Ito
formula to the condition $[X(t),Y(t)]=0$ for $Y(t)\in \{B(t,\mathrm{E}%
),B^{\ast }(t,\mathrm{E})\}$ we obtain 
\begin{eqnarray*}
\mathrm{d}[X(t),B(t,\mathrm{E})] &=&\int_{\mathrm{E}}([X(t),L_{x}(t)]-\delta
_{x}(t,X))\lambda (\mathrm{d}\mathbf{x})=0 \\
\mathrm{d}[X(t),B^{\ast }(t,\mathrm{E})] &=&\int_{\mathrm{E}%
}([X(t),L_{x}^{\ast }(t)]+\delta _{x}^{\ast }(t,X))\lambda (\mathrm{d}%
\mathbf{x})=0,
\end{eqnarray*}%
i. e. $\delta _{x}(X)=[X,L_{x}]$, $\delta _{x}^{\ast }(X)=[L_{x}^{\ast },X]$
due to $[\mathrm{d}X(t),Y(t)]=0$, $L_{x}(t)=j(t,L_{t,\mathbf{x}})$, $%
L_{x}^{\ast }(t)=j(t,L_{t,\mathbf{x}}^{\ast })$. This together with $\beta
(X)=j(i[H,X])$ gives $\gamma (t,X)=j(t,\gamma _{t}(X))$, where 
\begin{equation*}
\gamma _{t}(X)=i[H_{t},X]+\frac{1}{2}\ \int_{\Lambda }(L_{x}^{\ast
}[X,L_{x}]+[L_{x}^{\ast },X]L_{x})\lambda (\mathrm{d}\mathbf{x})
\end{equation*}%
is the solution of the equation 
\begin{equation*}
\gamma _{t}(X^{\ast }X)-X^{\ast }\gamma _{t}(X)-\gamma _{t}(X)^{\ast }X=\int
[L_{x}^{\ast },X][X,L_{x}]\lambda (\mathrm{d}\mathbf{x}),
\end{equation*}%
uniquelly defined up to a $\ast $-differentiation $\beta _{t}(X)=i[H_{t},X]$%
, $H_{t}=H_{t}^{\ast }$. The unique solution $j(t,X)=U_{t}XU_{t}^{\ast }$ of
the derived nonstationary Langevin equation under the boundness conditions (%
\ref{eq:ccr1.6}) was found in \cite{bib:bel5}, Corollary 4.

Let us denote by $U(s,t),\;s\geq t$ the solution of the quantum stochastic
evolution equation (\ref{eq:ccr1.5}) on the interval $(t,s]$ with $U(t,t)=I$
under the integrability conditions~(\ref{eq:ccr1.6}). The operators $U(s,r)$
commute with $\boldsymbol{Y}_{r}$, $r\leq s$, due to commutativity of $%
\boldsymbol{Y}_{r}\in \boldsymbol{A}(r),\boldsymbol{A}^{\ast }(r)$ and the
operators $\boldsymbol{L}_{t},\boldsymbol{L}_{t}^{\ast }$, $\mathrm{d}%
\boldsymbol{A}(t)$, $\mathrm{d}\boldsymbol{A}^{\ast }(t)$, $t\in \lbrack
r,s) $ generating $U(s,r)$. Hence $U_{s}\boldsymbol{Y}_{t}U_{s}^{\ast }=U_{t}%
\boldsymbol{Y}_{t}U_{t}^{\ast }$ because $U_{s}^{\ast }=U(s,t)^{\ast
}U_{t}^{\ast }$ for any $s>t$ and because of unitarity of $U(s,t)$. Using
the quantum Ito formula \cite{bib:12} one can easily find 
\begin{eqnarray*}
\mathrm{d}Y(t,\mathrm{E}) &=&\mathrm{d}(U_{t}Y_{t}(\mathrm{E})U_{t}^{\ast })=%
\mathrm{d}U_{t}Y_{t}(\mathrm{E})U_{t}^{\ast }+U_{t}\mathrm{d}Y_{t}(\mathrm{E}%
)U_{t}^{\ast }+U_{t}Y_{t}(\mathrm{E})\mathrm{d}U_{t}^{\ast } \\
&&+\mathrm{d}U_{t}\mathrm{d}Y_{t}(\mathrm{E})U_{t}^{\ast }+\mathrm{d}%
U_{t}Y_{t}(\mathrm{E})\mathrm{d}U_{t}^{\ast }+U_{t}\mathrm{d}Y_{t}(\mathrm{E}%
)\mathrm{d}U_{t}^{\ast }+\mathrm{d}U_{t}\mathrm{d}Y_{t}(\mathrm{E})\mathrm{d}%
U_{t}^{\ast }, \\
\mathrm{d}\boldsymbol{B}(t) &=&\mathrm{d}\boldsymbol{A}(t)+U_{t}\boldsymbol{L%
}_{t}U_{t}^{\ast }\mathrm{d}\lambda ,\;\;\mathrm{d}\boldsymbol{B}^{\ast }(t)=%
\mathrm{d}\boldsymbol{A}^{\ast }(t)+U_{t}\boldsymbol{L}_{t}^{\ast
}U_{t}^{\ast }\mathrm{d}\lambda , \\
\mathrm{d}Y(\mathrm{E}) &=&\mathrm{d}Q(\mathrm{E})+\int_{\mathrm{E}%
}U(L_{x}+L_{x}^{\ast })U^{\ast }\lambda (\mathrm{d}\mathbf{x})=\mathrm{d}Q(%
\mathrm{E})+M(\mathrm{E})\mathrm{d}t
\end{eqnarray*}%
for $Y(t,\mathrm{E})=B(t,\mathrm{E})+B^{\ast }(t,\mathrm{E})$, $\mathrm{E}%
\in \mathcal{B}$ due to the only nonzero infinitesimal multiplication $%
\mathrm{d}A(t,\mathrm{E}^{\prime })\mathrm{d}A^{\ast }(t,\mathrm{E})=\mathrm{%
d}t\lambda (\mathrm{E}\cap \mathrm{E}^{\prime })$, where $M$ is defined by (%
\ref{eq:ccr1.7}). The relation~(\ref{eq:ccr1.8}) for the process (\ref%
{eq:ccr1.7}) is a simple consequence of~(\ref{eq:ccr1.9}) and $[X,Q(\mathrm{E%
})]=0$ for any $\mathrm{E}\in \mathcal{A}(\Gamma _{1})$ and $X\in \mathcal{L}%
(\mathcal{H})\otimes I_{\mathcal{F}}$ : 
\begin{equation*}
\lbrack X(s),Y(t,\mathrm{E})]=[U_{s}XU_{s}^{\ast },\;U_{s}Y_{t}(\mathrm{E}%
)U_{s}^{\ast }]=U_{s}[X,Y_{t}(\mathrm{E})]U_{s}^{\ast }=0.
\end{equation*}

\textbf{Remark 1.} Considering instead of $\boldsymbol{Y}(t)=U_{s}%
\boldsymbol{Q}(t)U_{s}^{\ast }$ the sequential measurements of the output
momentum process $\boldsymbol{Y}(t)=U_{s}\boldsymbol{V}(t)U_{s}^{\ast }$, $%
s\geq t$, defined by $V(t,\mathrm{E})=\frac{1}{i}\ (A(t,\mathrm{E})-A^{\ast
}(t,\mathrm{E}))$ as 
\begin{equation*}
\mathrm{d}Y(t,\mathrm{E})=N(t,\mathrm{E})\mathrm{d}t+\mathrm{d}V(t,\mathrm{E}%
),\;\;\;\mathrm{E}\in \mathcal{B},
\end{equation*}%
where $N(t,\mathrm{E})=\frac{1}{i}\ \int_{\mathrm{E}}U(L_{t,\mathbf{x}}-L_{t,%
\mathbf{x}}^{\ast })U_{t}^{\ast }\lambda (\mathrm{d}\mathbf{x})$, one can
extract the information about the noncommuting self--adjoint operators $%
S_{x}=\frac{1}{i}\ (L_{x}-L_{x}^{\ast })$. Moreover, by doubling $\Lambda
\rightarrow \Lambda \times \{-,\,+\}$ the space $\Lambda $ and considering
the family $\{L_{t,\mathbf{x},-},\,L_{t,\mathbf{x},+}\}$ with $L_{t,\mathbf{x%
},\mp }=L_{t,\mathbf{x}}/\sqrt{\mp 2}$ instead of $\{L_{t,\mathbf{x}}\}$ one
can realize the continuous time--sequential indirect observation of the
pairs of operators 
\begin{equation*}
R_{x,+}=\frac{1}{\sqrt{2}}\ (L_{x}+L_{x}^{\ast }),\;\;R_{x,-}=\frac{1}{\sqrt{%
2}i}\ (L_{x}-L_{x}^{\ast }),\;\;x\in \mathbb{R}_{+}\times \Lambda
\end{equation*}%
by the measurement of the two commutative output processes $\boldsymbol{Y}%
_{\mp }(t)=U_{\infty }\boldsymbol{Q}_{\mp }(t)U_{\infty }^{\ast }$. Here $%
\boldsymbol{Q}_{\mp }(t)=\boldsymbol{A}_{\mp }(t)+\boldsymbol{A}_{\mp
}^{\ast }(t)$ are given by the independent Boson measures $A_{\mp }$ on $%
\mathcal{A}(\mathbb{R}_{+}\times \Lambda )$ as $A_{\mp }(t,\mathrm{E}%
)=A_{\mp }([0,t)\times \mathrm{E}),\;\mathrm{E}\in \mathcal{B}$, and $U_{t}$
satisfies the equation~(\ref{eq:ccr1.5}) with two-fold quantum stochastic
integral over $\Lambda \times \{-,\,+\}$ instead of $\Lambda $ which can be
written again as (\ref{eq:ccr1.5}) in terms of $A=\frac{1}{\sqrt{2}}\
(A_{+}+iA_{-})$. The complexified observable process $\boldsymbol{Z}=\frac{1%
}{\sqrt{2}}\ (\boldsymbol{Y}_{+}+i\boldsymbol{Y}_{-})$ defines the unsharp
observation $Z(t,\mathrm{E})=B(t,\mathrm{E})+A_{\circ }^{\ast }(t,\mathrm{E}%
) $, $\mathrm{E}\in \mathcal{B}$, $A_{\circ }^{\ast }=\frac{1}{\sqrt{2}}\
(A_{+}^{\ast }+iA_{-}^{\ast })$ of the nondemolition output field $%
\boldsymbol{B}(t)=U_{\infty }\boldsymbol{A}(t)U_{\infty }^{\ast }$.

In the case $\mathcal{B}=\mathcal{A}$ such the continuous measurement gives
a complete nondemolition sequential observation \cite{bib:b1} of the
non--Hermitian operators $L_{x}$ in terms of the complexified output process 
$\boldsymbol{Z}(t)=U_{\infty }\boldsymbol{W}(t)U_{\infty }^{\ast }$ having
the stochastic differential 
\begin{equation}
\mathrm{d}Z(t,\mathrm{E})=\mathrm{d}t\int_{\mathrm{E}}L_{\mathbf{x}%
}(t)\lambda (\mathrm{d}\mathbf{x})+\mathrm{d}W(t,\mathrm{E}),\;\;\mathrm{E}%
\in \mathcal{B},  \label{eq:ccr1.10}
\end{equation}%
where $L_{\mathbf{x}}(t)=U_{t}L_{t,\mathbf{x}}U_{t}^{\ast }$ and $%
\boldsymbol{W}(t)=\frac{1}{\sqrt{2}}\ (\boldsymbol{Q}_{+}(t)+i\boldsymbol{Q}%
_{-}(t))=\boldsymbol{A}(t)+\boldsymbol{A}_{\circ }^{\ast }(t)$ is the
complex Wiener process in Fock space over $L^{2}(\mathbb{R}_{+}\times
\Lambda \times \{-,+\})$ with multiplication table 
\begin{eqnarray*}
\mathrm{d}W^{\ast }(t,\mathrm{E})\mathrm{d}W(t,\mathrm{E}^{\prime }) &=&%
\mathrm{d}t\lambda (\mathrm{E}\cap \mathrm{E}^{\prime })=\mathrm{d}W(t,%
\mathrm{E}^{\prime })\mathrm{d}W^{\ast }(t,\mathrm{E}), \\
\mathrm{d}W(t,\mathrm{E})\mathrm{d}W(t,\mathrm{E}^{\prime }) &=&0,\;\;%
\mathrm{d}W^{\ast }(t,\mathrm{E})\mathrm{d}W^{\ast }(t,\mathrm{E}^{\prime
})=0.
\end{eqnarray*}

\section{A posteriori quantum dynamics under\newline
the continual measurements.}

Let us consider the quantum diffusion $j(t):\mathcal{L}\rightarrow \mathcal{L%
}(\mathcal{H}\otimes \mathcal{F})$ of the system over a unital $\ast $%
--algebra $\mathcal{L}$ in $\mathcal{H}$, together with the given
nondemolition output fields $\mathrm{d}\boldsymbol{B}=\boldsymbol{L}\mathrm{d%
}\lambda +\mathrm{d}\boldsymbol{A}$, $\mathrm{d}\boldsymbol{B}^{\ast }=%
\boldsymbol{L}^{\ast }\mathrm{d}\lambda +\mathrm{d}\boldsymbol{A}^{\ast }$.
The operators $j(t,X)=X(t)$ under the conditions of Theorem \ref{th:a1}
satisfy the quantum Langevin equation 
\begin{eqnarray}
&&\mathrm{d}X(t)-\mathrm{d}t\int_{\Lambda }\frac{1}{2}\ \left( L_{\mathbf{x}%
}^{\ast }(t)[X(t),\,L_{\mathbf{x}}(t)]+[L_{\mathbf{x}}^{\ast }(t),\,X(t)]L_{%
\mathbf{x}}(t)\right) \lambda (\mathrm{d}\mathbf{x})  \notag \\
&=&i[H(t),\,X(t)]\mathrm{d}t+\int_{\Lambda }\left( \mathrm{d}A^{\ast }(t,%
\mathrm{d}\mathbf{x})[X(t),\,L_{\mathbf{x}}(t)]+[L_{\mathbf{x}}^{\ast
}(t),\,X(t)]\mathrm{d}A(t,\mathrm{d}\mathbf{x})\right) ,  \label{eq:ccr2.1}
\end{eqnarray}%
having the unique solution $X(t)=U_{t}XU_{t}^{\ast }$, where $U_{t}^{\ast }$%
, $t\in \mathbb{R}_{+}$ are the unitary operators defined by the QS equation
(\ref{eq:ccr1.5}), and 
\begin{equation*}
K(t)=U_{t}K_{t}U_{t}^{\ast },\;\;K^{\ast }(t)=U_{t}K_{t}^{\ast }U_{t}^{\ast
},\;\;L_{\mathbf{x}}(t)=U_{t}L_{\mathbf{x},t}U_{t}^{\ast },\;\;L_{\mathbf{x}%
}^{\ast }(t)=U_{t}L_{\mathbf{x},t}^{\ast }U_{t}^{\ast }.
\end{equation*}%
The equation (\ref{eq:ccr2.1}) can be obtained from~(\ref{eq:ccr1.5}) by
using the QS Ito formula 
\begin{equation*}
\mathrm{d}(U_{t}XU_{t}^{\ast })=\mathrm{d}U_{t}XU_{t}^{\ast }+U_{t}X\mathrm{d%
}U_{t}^{\ast }+\mathrm{d}U_{t}X\mathrm{d}U_{t}^{\ast }
\end{equation*}%
and the Hudson--Parthasarathy multiplication table \cite{bib:12} 
\begin{eqnarray*}
\mathrm{d}A^{\ast }(t,\mathrm{E})\mathrm{d}A(t,\mathrm{E}^{\prime })
&=&0,\;\;\mathrm{d}A(t,\mathrm{E}^{\prime })\mathrm{d}A^{\ast }(t,\mathrm{E}%
)=\mathrm{d}t\lambda (\mathrm{E}\cap \mathrm{E}^{\prime }), \\
\mathrm{d}A(t,\mathrm{E})\mathrm{d}A(t,\mathrm{E}^{\prime }) &=&0,\;\mathrm{d%
}A^{\ast }(t,\mathrm{E})\mathrm{d}A^{\ast }(t,\mathrm{E}^{\prime
})=0,\;\forall \mathrm{E},\mathrm{E}^{\prime }\in \mathcal{A}
\end{eqnarray*}%
The \emph{a posterior dynamics} of the system under the observation (\ref%
{eq:ccr1.7}) with a given initial state $\phi _{0}$ is the dynamics $\phi
_{0}\mapsto \hat{\pi}_{t}$, $t\in \mathbb{R}_{+}$ of the a posterior state $%
\hat{\pi}_{t}$ on $\mathcal{L}$, giving posterior mean values $\hat{x}_{t}=%
\hat{\pi}_{t}\{X\}$ of $X\in \mathcal{L}$ as stochastic functions of the
trajectories of the observed process $\boldsymbol{Y}^{t}=\{\boldsymbol{Y}%
(r)|r\leq t\}$. According to \cite{bib:2} the \emph{a posterior state} is
defined by the conditional expectation $\epsilon \{X\}(t)=\epsilon
_{t}\{X(t)|\boldsymbol{Y}^{t}\}$ on the commutant $\mathcal{N}_{t}=\{%
\boldsymbol{Y}(r)|r\leq t\}^{\prime }$ in $\mathcal{L}(\mathcal{H}\otimes 
\mathcal{F})$, which contains $\boldsymbol{Y}^{t}$ and $X(t)$ due to the
nondemolition property~(\ref{eq:ccr1.8}). By Theorem~\ref{th:a1} the
operators $\epsilon \{X\}(t)\in \mathcal{N}_{t}^{\prime }$ have in the Schr%
\"{o}dinger picture the form 
\begin{equation}
U_{t}^{\ast }\epsilon \{X\}(t)U_{t}=U_{t}^{\ast }\epsilon
_{t}\{U_{t}XU_{t}^{\ast }|\boldsymbol{Y}^{t}\}U_{t}=I\otimes \hat{\pi}%
_{t}\{X\},\;\;\forall X\in \mathcal{L}  \label{eq:ccr2.2}
\end{equation}%
since $U_{t}^{\ast }\mathcal{N}_{t}^{\prime }U_{t}$ commutes with $\mathcal{L%
}(\mathcal{H})\otimes I$. As a map $\hat{\pi}_{t}:\;\mathcal{L\rightarrow M}%
_{t}$ into the Abelian algebra $\mathcal{M}_{t}=U_{t}^{\ast }\mathcal{N}%
_{t}^{\prime }U_{t}\subset \mathcal{L}(\mathcal{F})$ generated by $\{%
\boldsymbol{Y}_{r}|r\leq t\}$ on $\mathcal{F}$, the a posterior state
satisfies a nonlinear stochastic equation , obtained for the first time with
respect to $\boldsymbol{Y}(t)$ as the quantum filtering equation in \cite%
{bib:2,bib:b7}. Here we shall derive a linear quantum stochastic equation
for a nonnormalized posterior state $\hat{\phi}_{t}\{X\}=\hat{\rho}_{t}\hat{%
\pi}_{t}\{X\}$, where $\hat{\rho}_{t}$ is a positive stochastic functional $%
\hat{\rho}_{t}=\hat{\rho}(\boldsymbol{Y}^{t})$ of $\boldsymbol{Y}_{r}=%
\boldsymbol{Q}(r)$, $r\leq t$. Moreover, we shall prove that the stochastic
normalization factor $\hat{\rho}_{t}$ can be taken as the probability
density $\hat{\rho}(\boldsymbol{v}^{t})$ of the trajectories $\boldsymbol{v}%
^{t}=\{\boldsymbol{v}(r)|r\leq t\}$ of the observed process $\boldsymbol{Y}%
^{t}$ with respect to the standard probability measure $\nu $ of a Wiener
process $\boldsymbol{w}$, represented in the Fock space $\mathcal{F}$ as $%
\boldsymbol{Q}$ with respect to the vacuum state $|0\rangle \in \mathcal{F}$%
. Once the density operator $\hat{\rho}_{t}=\hat{\phi}_{t}\{I\}$ is found by
the solution of the linear posterior evolution equation, the density
function $\hat{\rho}(\boldsymbol{v}^{t})=\rho _{t}^{\boldsymbol{w}}$ is
given by the Segal (duality) transformation $\boldsymbol{Q}\mapsto 
\boldsymbol{w}$ of the observable process $\boldsymbol{Q}^{t}=U_{t}^{\ast }%
\boldsymbol{Y}^{t}U_{t}$ in the Schr\"{o}dinger picture.

We shall say the nondemolition observation is \emph{complete} for the
quantum diffusion, described by the stochastic evolution equation (\ref%
{eq:ccr2.1}), if the subsets $\mathrm{E}\in \mathcal{B}$ in (\ref{eq:ccr1.7}%
) generate the $\sigma $-algebra $\mathcal{A}$. Let us see now in that case
the posterior dynamics is not mixing: $\hat{\pi}_{t}=\hat{T}_{t}\phi _{0}%
\hat{T}_{t}^{\ast }$, i. e. it is defined as $\phi _{t}^{\boldsymbol{w}%
}\{X\}=(\varphi _{t}^{\boldsymbol{w}}|X\varphi _{t}^{\boldsymbol{w}})$, for $%
\phi _{0}\{X\}=\left( \psi |X\psi \right) $ by a posterior stochastic
propagator $T_{t}^{\boldsymbol{w}}:\psi \in \mathcal{H}\mapsto \varphi _{t}^{%
\boldsymbol{w}}=T_{t}^{\boldsymbol{w}}\psi $. We show the renormalized
propagator $\hat{F}_{t}^{\boldsymbol{w}}=\sqrt{\rho _{t}^{\boldsymbol{w}}}%
T_{t}^{\boldsymbol{w}}$ also satisfies a linear stochastic wave equation $%
\hat{F}+K\hat{F}\mathrm{d}t=\boldsymbol{L}\mathrm{d}\boldsymbol{w}\hat{F}$
in $\mathcal{H}$, given in the Fock space representation by the operator
evolution equation in $\mathcal{H}\otimes \mathcal{F}$,%
\begin{equation}
\mathrm{d}\hat{F}_{t}+K_{t}\hat{F}_{t}\mathrm{d}t=\int_{\Lambda }L_{t,%
\mathbf{x}}\hat{F}_{t}\mathrm{d}Y_{t}(\mathrm{d}\mathbf{x}),\;\;\hat{F}%
_{0}=I,  \label{eq:ccr2.3}
\end{equation}%
where $\boldsymbol{L}_{t}\mathrm{d}\boldsymbol{Y}_{t}:=\int_{\Lambda }L_{t,%
\mathbf{x}}\mathrm{d}Y_{t}(\mathrm{d}\mathbf{x})=\boldsymbol{L}_{t}\mathrm{d}%
\boldsymbol{Q}(t)$, $(\hat{F}_{t}(\boldsymbol{w})\psi \mid \hat{F}_{t}(%
\boldsymbol{w})\psi )=\rho _{t}^{\boldsymbol{w}}$. The proof is given in
Lemma~\ref{lem:a1} and Lemma~\ref{lem:a2} in terms of $\hat{\Phi}_{t}\{X\}=%
\hat{F}_{t}^{\ast }X\hat{F}_{t}$.

Firstly let us note that the wave propagator $\hat{F}_{t}:\;\mathcal{H}%
\rightarrow \mathcal{H}\otimes \mathcal{M}_{t}$ as any other adapted Wiener
functional of $Y$ is defined in the Fock representation $F^{t}=\hat{F}%
_{t}|0\rangle $ by the generating functional $F_{g}^{t}=\int_{\Gamma
}F^{t}(\chi )\prod_{x\in \chi }g(x)\lambda (\mathrm{d}x)$ coinsiding with
the Wick symbol $\langle f|\hat{F}_{t}|f\rangle =F_{g}^{t}$ for $g=f+f^{\ast
}$, where $|f\rangle \in \mathcal{F}$ is the coherent state 
\begin{equation*}
|f\rangle (\chi )=e^{-\Vert f\Vert ^{2}/2}\prod_{x\in \chi }f(x),\;\Vert
f\Vert ^{2}=\int |f(x)|^{2}\lambda (\mathrm{d}x)
\end{equation*}%
for a $f:\;\Gamma _{1}\rightarrow \mathbb{C}$ with $\Vert f\Vert ^{2}<\infty 
$, denoted as $f^{2}=\Vert f\Vert ^{2}$, if $f^{\ast }=f$. It helps to prove
the

\begin{lemma}
The solution $\hat{F}_t$ of the stochastic equation (\ref{eq:ccr2.3})
satisfies the equivalency condition $\hat{F}_{t}| 0\rangle= U_{t}^{\ast}|
0\rangle$ respectively to the vacuum $|0\rangle\in\mathcal{F}$ with the
unitary propagator $U_{t}^{\ast}$ defined by the equation (\ref{eq:ccr1.5}),
i.e. $\hat{F}_{t}h=U_{t}^{\ast}h$ for all $h=\psi\otimes| 0\rangle$, where $%
\psi\in\mathcal{H}$, $t\geq 0$.\label{lem:a1}
\end{lemma}

\textbf{Proof}. To this end we note that $\mathcal{A}$-measurability
coincides with $\mathcal{B}$-measurability in this case and the equation for 
$F_{t}^{\ast }=U_{t}^{\ast }|0\rangle $ with $F_{0}^{\ast }=I$ can be simply
obtained by allowing the right hand side of (\TEXTsymbol{\backslash}%
ref\{eq:ccr1.5\}) to act on the Fock vacuum $|0\rangle $. This gives%
\begin{equation*}
(\boldsymbol{L}_{t}\mathrm{d}\boldsymbol{A}^{\ast }(t)-\boldsymbol{L}%
_{t}^{\ast }\mathrm{d}\boldsymbol{A}(t))U_{t}^{\ast }|0\rangle =(\boldsymbol{%
L}_{t}\mathrm{d}\boldsymbol{A}^{\ast }(t)+\boldsymbol{L}_{t}\mathrm{d}%
\boldsymbol{A}(t))U_{t}^{\ast }|0\rangle =\boldsymbol{L}_{t}\mathrm{d}%
\boldsymbol{Y}_{t}F_{t}^{\ast },
\end{equation*}%
where $\boldsymbol{L}_{t}\mathrm{d}\boldsymbol{A}^{\ast }(t)=\int_{\Lambda
}L_{t,\mathbf{x}}\mathrm{d}A^{\ast }(t,\mathrm{d}\mathbf{x})$ due to $%
\mathcal{B}$-measurability of the map $\boldsymbol{L}_{t}:\mathbf{x}\mapsto
L_{t,\mathbf{x}}$ for almost all $t$, the commutativity of the increments $%
\mathrm{d}B_{t}(\mathrm{E})=\mathrm{d}A(t,\mathrm{E})$ with $U_{t}^{\ast }$
and with $\hat{F}_{t}$ and the annihilation property $\mathrm{d}B_{t}(%
\mathrm{E})|0\rangle =0=\mathrm{d}A(t,\mathrm{E})|0\rangle $ for all $%
\mathrm{E}\in \mathcal{A}$. The equation for the $\mathcal{L(H)}$-valued
symbol $F_{g}^{t}=\langle g|\hat{F}_{t}|0\rangle e^{g^{2}/2}$ of the
nonunitary classical stochastic evolution $\hat{F}_{t}$ defined by $(h_{g}|%
\hat{F}_{t}h_{0}^{\prime })=(\psi |F_{g}^{t}\psi ^{\prime })$ for all $%
h_{g}=\psi \otimes e^{g^{2}/2}|g\rangle $, $h_{0}^{\prime }=\psi ^{\prime
}\otimes |0\rangle $, is given by 
\begin{equation}
\frac{\mathrm{d}}{\mathrm{d}t}\ F_{g}^{t}+K_{t}F_{g}^{t}=\int_{\Lambda }L_{t,%
\mathbf{x}}F_{g}^{t}g(t,\mathbf{x})\lambda (\mathrm{d}\mathbf{x}%
),\;\;g=g^{\ast }\in L^{2}(\Gamma _{1}).  \label{eq:ccr2.4}
\end{equation}%
This coinsides with the equation for $\langle g|U_{t}^{\ast }|0\rangle
e^{g^{2}/2}=F_{g,t}^{\ast }$ having the same form as~(\ref{eq:ccr2.3}) with $%
F_{g,t}^{\ast }$ instead of $\hat{F}_{t}$ and $\boldsymbol{G}%
_{t}=\int_{0}^{t}\boldsymbol{g}(r)\mathrm{d}r$, $\boldsymbol{g}(t)=%
\boldsymbol{f}(t)+\boldsymbol{f}^{\ast }(t)$ instead of $\boldsymbol{Y}_{t}=%
\boldsymbol{B}_{t}+\boldsymbol{B}_{t}^{\ast }$, $B_{t}(\mathrm{E})=A(t,%
\mathrm{E})$, $\mathrm{E}\in \mathcal{B}$, and the initial operator $%
F_{g,0}^{\ast }=I$. It means that $F_{g,t}^{\ast }=F_{t}^{g}$ and $%
U_{t}^{\ast }|0\rangle =\hat{F}_{t}|0\rangle $ due to the uniqueness of the
solution of the equation (\ref{eq:ccr2.3}) proved in \cite{bib:bel5} under
the conditions~(\ref{eq:ccr1.6}).\hfill $\Box $

Secondly, let us find the QS Langevin equation for the process $%
X_{g}(t)=U_{t}X_{g}^{t}U_{t}^{\ast }$, $X_{g}^{t}=\hat{e}_{g}^{t}X$, where $%
g\in L_{\mathcal{B}}^{2}(\Gamma _{1})$ is a $\mathcal{B}(\mathbb{R}%
_{+})\otimes \mathcal{B}$--measurable square integrable function and $\hat{e}%
_{g}^{t}=:e^{q(g_{t})}:=\hat{e}_{g_{t}}$ is the Wick ordered exponential 
\begin{equation*}
\hat{e}_{g_{t}}=:\!\exp \int_{0}^{t}\int_{\Lambda }g(r,\mathbf{x})\mathrm{d}%
Q(r,\mathrm{d}\mathbf{x})\!:=e^{a^{\ast }(g_{t})}e^{a(g_{t})}
\end{equation*}%
of the observable $y_{t}(g)=\int_{0}^{t}\boldsymbol{g}\mathrm{d}\boldsymbol{Y%
}=q(g_{t})$ in the Schr\"{o}dinger picture, corresponding to the product $%
e_{g}(\chi )=\prod_{x\in \chi }g(x)$, $\chi \in \Gamma $ in the Fock
representation of $e_{g}=\hat{e}_{g}|0\rangle $. Here and below $g_{t}\in L_{%
\mathcal{B}}^{2}(\Gamma _{1})$ denotes the projection of $g$ with $%
g_{t}(x)=0 $, if $x\notin \lbrack 0,t)\times \mathrm{E}$ for every $\mathrm{E%
}\in \mathcal{B}$, otherwise $g_{t}=g$, and $a^{\ast }(g_{t})=\int
g_{t}(x)A^{\ast }(\mathrm{d}x)=a(g_{t}^{\ast })^{\ast }$, $%
q(g_{t})=(a+a^{\ast })(g_{t})$. Taking into account that this exponential is
defined by the equation $\mathrm{d}\hat{e}_{g}^{t}=\hat{e}_{g}^{t}%
\boldsymbol{g}(t)\mathrm{d}\boldsymbol{Y}_{t}$ with $\hat{e}_{0}=1$, we can
obtain for $G(t)=U_{t}X_{g}^{t}U_{t}^{\ast }=\hat{e}_{g}(t)X(t)$ 
\begin{eqnarray*}
\mathrm{d}G+(GK+K^{\ast }G)\mathrm{d}t &=&\mathrm{d}t\int_{\Lambda }\{L_{%
\mathbf{x}}^{\ast }GL_{\mathbf{x}}+(L_{\mathbf{x}}^{\ast }G+GL_{\mathbf{x}%
})g(\mathbf{x})\}\lambda (\mathrm{d}\mathbf{x}) \\
&&+\int_{\Lambda }\{[L_{\mathbf{x}}^{\ast },G]\mathrm{d}B(\mathrm{d}\mathbf{x%
})+\mathrm{d}B^{\ast }(\mathrm{d}\mathbf{x})[G,L_{\mathbf{x}}] \\
&&+g(\mathbf{x})(\mathrm{d}B(\mathrm{d}\mathbf{x})+\mathrm{d}B^{\ast }(%
\mathrm{d}\mathbf{x}))G\}
\end{eqnarray*}%
using the quantum Ito formula $\mathrm{d}(\hat{e}X)=\mathrm{d}\hat{e}X+\hat{e%
}\mathrm{d}X+\mathrm{d}\hat{e}\mathrm{d}X$ for $\hat{e}_{g}(t)=U_{t}\hat{e}%
_{g}^{t}U_{t}^{\ast }$ and~(\ref{eq:ccr2.1}). It helps to write the equation
for the vacuum expectation operator 
\begin{equation*}
\Phi _{g}^{t}\{X\}=\langle 0|G(t)|0\rangle =F_{t}\hat{e}_{g}^{t}XF_{t}^{\ast
},\;\;\Phi _{g}^{t}\{I\}=\mathrm{P}_{g_{t}}:=\Phi
_{g_{t}}^{s}\{I\},\;\forall s\geq t
\end{equation*}%
as $\langle 0|\{\mathrm{d}G+(GK+K^{\ast }G)\mathrm{d}t\}(t)|0\rangle =%
\mathrm{d}t\langle 0|\{\boldsymbol{L}^{\ast }G\boldsymbol{L}+(\boldsymbol{L}%
^{\ast }G+G\boldsymbol{L})\boldsymbol{g}\}(t)|0\rangle $, or equivalently 
\begin{eqnarray}
&&\frac{\mathrm{d}}{\mathrm{d}t}\ \Phi _{g}^{t}\{X\}+\Phi
_{g}^{t}\{K_{t}^{\ast }X+XK_{t}\}  \notag \\
&=&\int_{\Lambda }\Phi _{g}^{t}\{L_{t,\mathbf{x}}^{\ast }XL_{t,\mathbf{x}%
}+(XL_{t,\mathbf{x}}+L_{t,\mathbf{x}}^{\ast }X)g(t,\mathbf{x})\}\lambda (%
\mathrm{d}\mathbf{x}).  \label{eq:ccr2.5}
\end{eqnarray}%
The equation (\ref{eq:ccr2.5}), with $\Phi _{g}^{0}\{X\}=X$, defines both
the prior quantum Markovian dynamics \cite{bib:bel4} $\mathrm{M}%
^{t}:X\mapsto F_{t}XF_{t}^{\ast }$ as $\mathrm{M}^{t}=\Phi _{0}^{t}$ and an
operator--valued generating functional $\mathrm{P}_{g}=F_{\infty }\hat{e}%
_{g}F_{\infty }^{\ast }=\lim_{t\rightarrow \infty }\Phi _{g}^{t}\{I\}$ of
factorial (normal ordered) moment operators 
\begin{equation*}
\langle 0|:\!\dot{Y}(x_{1})\ldots \dot{Y}(x_{n})\!:|0\rangle =\delta ^{n}%
\mathrm{P}_{g}/\delta g(x_{1})\ldots \delta g(x_{n})\mid _{g=0}
\end{equation*}%
for the measurements at $t_{m}<t,\;\mathbf{x}_{m}\in \Lambda ,\;m=1,\ldots
,n $ of generalized derivatives $\dot{Y}(x)=Y(\mathrm{d}x)/\lambda (\mathrm{d%
}x)\equiv \dot{Y}_{\mathbf{x}}(t)$ of the measure $\boldsymbol{Y}(\mathrm{d}%
x) $ on $\mathcal{B}(\mathbb{R}_{+})\otimes \mathcal{B}$. It follows from
the Weyl representation 
\begin{equation}
\hat{e}_{g}^{t}(q)=\exp \int_{0}^{t}\{\int_{\Lambda }g(r,\mathbf{x})\mathrm{d%
}Y_{r}(\mathrm{d}\mathbf{x})-\frac{1}{2}\ \int_{\Lambda }g(r,\mathbf{x}%
)^{2}\lambda (\mathrm{d}\mathbf{x})\mathrm{d}r\}=e^{q(g_{t})-g_{t}^{2}/2}
\label{eq:ccr2.6}
\end{equation}%
of the Wick exponent $\hat{e}_{g}^{t}=:\!e^{q(g_{t})}\!:$, that equation (%
\ref{eq:ccr2.5}) defines the characteristic operator $\Theta
_{g}^{t}\{X\}=\langle 0|e^{iy(g_{t})}X(t)|0\rangle $ of $y(g_{t})=\int
g_{t}(x)Y(\mathrm{d}x)=U_{t}q_{t}(g_{t})U_{t}^{\ast }$: 
\begin{equation*}
\Theta _{g}^{t}\{X\}=F_{t}e^{iq_{t}(g_{t})}XF_{t}^{\ast
}=e^{-g_{t}^{2}/2}\Phi _{ig}^{t}\{X\}.
\end{equation*}

Let us denote by $\boldsymbol{v}_{t}=\{v_{t}(\mathrm{E})|\mathrm{E}\in 
\mathcal{B}\}$ a stochastic trajectory $v_{t}(\mathrm{E}):\Omega \rightarrow 
\mathbb{R}$ of the process $\boldsymbol{Y}_{t}$ in the Wiener representation 
$v_{t}(\mathrm{E})=Y_{t}(\mathrm{E},\omega )$ and by $\upsilon
_{t}(g)=\int_{0}^{t}\boldsymbol{g}\mathrm{d}\boldsymbol{v}$, the Wiener
integral of $g(x),\;x\in \Gamma _{1}$. Now we prove the absolute continuity $%
\mathrm{I}_{t}\{X\}(\mathrm{d}\omega )=\hat{\Phi}_{t}\{X\}(\upsilon _{t})%
\mathrm{d}\mu (\omega )$ of the corresponding instrument $\mathrm{I}%
_{t}\{X\}(\mathrm{E})$, $\mathrm{E}\in \mathcal{B}$ with respect to the
standard Wiener measure $\mathrm{d}\mu (\omega )$.

\begin{lemma}
The solution of the equation (\ref{eq:ccr2.5}) is given by the expectation 
\begin{equation}
\Phi _{g}^{t}\{X\}=\int_{\Omega }e^{\upsilon _{t}(g)-g_{t}^{2}/2}\hat{\Phi}%
_{t}\{X\}(\upsilon _{t})\mathrm{d}\mu (\omega )  \label{eq:ccr2.7}
\end{equation}%
of a stochastic map $\Phi _{t}^{\omega }:\;X\mapsto \hat{\Phi}%
_{t}\{X\}(\upsilon _{t})$ as the nonanticipating function $\Phi _{t}^{\omega
}=\hat{\Phi}_{t}(\upsilon _{t})$ of $\boldsymbol{v}_{r}$, $r<t$, normalized
by a stochastic operator --function $\mathrm{P}_{t}^{\omega }=\hat{\mathrm{P}%
}(\upsilon _{t})$ and the factorial exponent~(\ref{eq:ccr2.6}) of the
representation $q(g_{t})\mapsto \upsilon _{t}(g)$. \label{lem:a2}
\end{lemma}

\textbf{Proof.} Let us take $\Omega $ as the spectrum of the commutative
field measure $Q(\mathrm{d}x)$, denoted as $w(\mathrm{d}x)$ in the standard
Wiener representation $\omega (f)=\int f(x)w(\mathrm{d}x),\;\;f\in
L^{2}(\Gamma _{1})$ and $\mu $ as the Gaussian probability measure on $%
\Omega $ with the correlations 
\begin{equation*}
\int_{\Omega }w(\Delta )w(\Delta ^{\prime })\mathrm{d}\mu (\omega )=\lambda
(\Delta \cap \Delta ^{\prime })=\langle 0|Q(\Delta )Q(\Delta ^{\prime
})|0\rangle
\end{equation*}%
induced by the Fock vacuum state. Then $\omega (g_{t})=\upsilon _{t}(g)$ as $%
q(g_{t})=y_{t}(g)$ for every $\mathcal{B}$-measurable $g:\Gamma
_{1}\rightarrow \mathbb{R}$, and due to $U_{t}^{\ast }|0\rangle =\hat{F}%
_{t}|0\rangle $ and the commutativity $\hat{e}_{g}^{t}\hat{F}_{t}=\hat{F}_{t}%
\hat{e}_{g}^{t}$ one can obtain 
\begin{eqnarray*}
\Phi _{g}^{t}\{X\} &=&\langle 0|U_{t}\hat{e}_{g}^{t}XU_{t}^{\ast }|0\rangle
=\langle 0|\hat{e}_{g}^{t}\hat{F}_{t}^{\ast }X\hat{F}_{t}|0\rangle \\
&=&\int_{\Omega }\hat{e}_{g}^{t}(\omega )\hat{F}_{t}^{\ast }(\omega )X\hat{F}%
_{t}(\omega )\mathrm{d}\mu (\omega ) \\
&=&\int_{\Omega }e^{\upsilon _{t}(g)-g_{t}^{2}/2}\hat{F}_{t}^{\ast }(\omega
)X\hat{F}_{t}(\omega )\mathrm{d}\mu (\omega ).
\end{eqnarray*}%
Here $\hat{F}_{t}(\omega )=F_{t}^{\omega }$ is the solution $\hat{F}%
_{t}=F_{t}^{q}$ of the equation~(\ref{eq:ccr2.3}) as the functional of $%
Y_{r}(\mathrm{E})=q(1_{r}(\mathrm{E}))$, $r<t$, $\mathrm{E}\in \mathcal{B}$
in the Wiener representation, where $1_{r}(\mathrm{E})$ is the indicator of $%
[0,r)\times \mathrm{E}$, and $\hat{e}_{g}^{t}(\omega )=e^{\upsilon
_{t}(g)-g_{t}^{2}/2}$ is the Wick exponent~(\ref{eq:ccr2.6}). Due to
arbitrariness of $\mathcal{B}(\mathbb{R}_{+})\otimes \mathcal{B}$%
--measurable $g$, it defines the posterior map $\hat{\Phi}_{t}=\Phi
_{t}(y_{t})$ in~(\ref{eq:ccr2.7}) as the classical conditional expectation 
\begin{equation}
\hat{\Phi}_{t}\{X\}(\upsilon _{t})=\int_{\Omega }\hat{F}_{t}^{\ast }(\omega
)X\hat{F}_{t}(\omega )\mathrm{d}\mu (\omega |\upsilon _{t})
\label{eq:ccr2.8}
\end{equation}%
with respect to the $\sigma $--algebra on $\Omega $, generated by the data $%
v_{r}(\mathrm{E})=w([0,r)\times \mathrm{E})$, $r\in \lbrack 0,t)$, $\mathrm{E%
}\in \mathcal{B}$. It is given by integrating on $\Omega $ with the Gaussian
conditional measure $\mathrm{d}\mu (\omega |\upsilon _{t})=\mathrm{d}\mu
(\omega )/\mathrm{d}\mu (\upsilon _{t})$, where $\mathrm{d}\mu (\upsilon
_{t})$ is the induced Gaussian probability measure on the trajectories $%
\boldsymbol{v}^{t}=\{\boldsymbol{v}(r)|r<t\}=\boldsymbol{w}^{t}|\mathcal{B}$
of the standard Wiener measure $\boldsymbol{w}(t,\mathrm{E})=\boldsymbol{w}%
([0,t)\times \mathrm{E})$ on $\mathcal{B}\ni \mathrm{E}$. Hence the
probability measure of the data $\boldsymbol{v}^{t}$ for the nondemolition
observation (\ref{eq:ccr1.7}) with a given initial wave function $\psi \in 
\mathcal{H}$ has the density 
\begin{equation*}
\rho _{t}^{\omega }=\int \Vert F_{t}(\omega )\psi \Vert ^{2}\mathrm{d}\mu
(\omega |\upsilon _{t})=(\psi |\hat{\mathrm{P}}(\upsilon _{t})\psi )\equiv 
\hat{\rho}(\upsilon _{t}),
\end{equation*}%
where $\hat{\mathrm{P}}(\upsilon _{t})=\hat{\Phi}_{t}\{I\}(\upsilon )=%
\mathrm{P}_{t}^{\omega }$. The non-Gaussian measure $\mathrm{d}\nu =\rho 
\mathrm{d}\mu $ defines the factorial generating functionals $\rho
_{g}^{t}=\langle \hat{e}_{g}^{t}(y)\rangle $ for the process $\boldsymbol{Y}%
^{t}$ as $(\psi |\Phi _{g}^{t}\{I\}\psi )$ and the mean values $\langle
X(t)\rangle $ of the operators $X(t)$ at the initial states $\psi \in 
\mathcal{H}$ as $(\psi |\Phi _{t}^{(0)}\{X\}\psi )$ by the averaging 
\begin{equation*}
(\psi |\Phi _{g}^{t}\{X\}\psi )=\int e^{\upsilon _{t}(g)-g_{t}^{2}/2}\hat{\pi%
}_{t}\{X\}(\upsilon _{t})\mathrm{d}\nu (\upsilon _{t})=\phi _{g}^{t}\{X\}
\end{equation*}%
of the product $\hat{e}_{g}^{t}(\upsilon _{t})\hat{\pi}_{t}\{X\}(\upsilon
_{t})$, where $\hat{\pi}_{t}\{X\}(\upsilon _{t})=(\psi |\hat{\Phi}%
_{t}\{X\}(\upsilon _{t})\psi )/\hat{\rho}(\upsilon _{t})$, over all the
observed in the past trajectories $\boldsymbol{v}^{t}$.\hfill $\Box $

Let us derive the corresponding linear stochastic equation for the
non--normalized posterior map (\ref{eq:ccr2.8}) $X\mapsto \hat{\Phi}%
_{t}\{X\} $ defining the posterior transformation $\phi _{0}\mapsto \phi
_{0}\circ \hat{\Pi}$ for any initial $\phi _{0}$ by $\hat{\Pi}_{t}\{X\}=\hat{%
\Phi}_{t}\{X\}/\hat{\rho}_{t}$, $\hat{\rho}_{t}=\phi _{0}\{\hat{\mathrm{P}}%
_{t}\}$. In the case of a complete nondemolition observation it can be
obtained in the Schr\"{o}dinger picture from~(\ref{eq:ccr2.3}) in the same
way as~(\ref{eq:ccr2.1}) from ~(\ref{eq:ccr1.5}) by using the Ito's formula
for $\hat{F}_{t}^{\ast }X\hat{F}_{t}=\hat{\Phi}_{t}\{X\}$: 
\begin{eqnarray*}
&&\mathrm{d}(\hat{F}_{t}^{\ast }X\hat{F}_{t})+\hat{F}_{t}(XK_{t}+K_{t}^{\ast
}X-\int_{\Lambda }L_{t,\mathbf{x}}^{\ast }XL_{t,\mathbf{x}}\lambda (\mathrm{d%
}\mathbf{x}))\hat{F}_{t}\mathrm{d}t \\
&=&\int_{\Lambda }\hat{F}_{t}^{\ast }(XL_{t,\mathbf{x}}+L_{t,\mathbf{x}%
}^{\ast }X)\hat{F}_{t}\mathrm{d}Y_{t}(\mathrm{d}\mathbf{x}).
\end{eqnarray*}%
In the general case the stochastic differential equation for~(\ref{eq:ccr2.8}%
) gives the following theorem.

\begin{theorem}
The conditional expectation ~(\ref{eq:ccr2.8}), defining in~(\ref{eq:ccr2.7}%
) the absolutely continuous operational measure $\hat{\Phi}%
_{t}\{X\}(\upsilon _{t})\mathrm{d}\mu (\omega )$ with respect to the Wiener
process $v_{t}(\omega )$, represented in Fock space by $\boldsymbol{Y}%
_{t}=\{Y_{t}(\mathrm{E})|\mathrm{E}\in \mathcal{B}\}$, satisfies the linear
stochastic equation 
\begin{eqnarray}
&&\mathrm{d}\hat{\Phi}_{t}\{X\}+\hat{\Phi}_{t}\{XK_{t}+K_{t}^{\ast
}X-\int_{\Lambda }L_{t,\mathbf{x}}^{\ast }XL_{t,\mathbf{x}}\lambda (\mathrm{d%
}\mathbf{x})\}\mathrm{d}t  \notag \\
&=&\int_{\Lambda }\hat{\Phi}_{t}\{X\bar{L}_{t,\mathbf{x}}+\bar{L}_{t,\mathbf{%
x}}^{\ast }X\}\mathrm{d}Y_{t}(\mathrm{d}\mathbf{x})  \label{eq:ccr2.9}
\end{eqnarray}%
corresponding to the equation (\ref{eq:ccr2.5}) for the Wick symbol $\Phi
_{g}^{t}\{X\}=\langle f|\hat{\Phi}_{t}|f\rangle $, $g=2\Re \bar{f}_{t}$.
Here $\bar{L}_{t,\mathbf{x}}$ are $\mathcal{B}$--measurable operator--valued
functions of $\mathbf{x}\in \Lambda $, $\bar{L}_{t,\mathbf{x}}=0$, if $%
\mathbf{x}\notin \mathrm{E}$ for any $\mathrm{E}\in \mathcal{B}$, defined
for almost all $t$ as a conditional averaging of $L_{t,\mathbf{x}}$ with
respect to $\mathcal{B}\subseteq \mathcal{A}$ by 
\begin{equation*}
\int_{\Lambda }\bar{L}_{t,\mathbf{x}}g(\mathbf{x})\lambda (\mathrm{d}\mathbf{%
x})=\int_{\Lambda }L_{t,\mathbf{x}}g(\mathbf{x})\lambda (\mathrm{d}\mathbf{x}%
)
\end{equation*}%
for any $\mathcal{B}$--measurable square--integrable $g:\Lambda \mapsto 
\mathbb{R}$ and $\bar{f}(t,\mathbf{x})$ is defined similary by the averaging
of $f(t,\mathbf{x})$. In particular, $\bar{L}_{t,\mathbf{x}}=\frac{1}{%
\lambda (\mathrm{E}_{i})}\ \int_{\mathrm{E}_{i}}L_{t,\mathbf{x}}\lambda (%
\mathrm{d}\mathbf{x})$ for all $\mathbf{x}\in \mathrm{E}_{i}$, if $\mathcal{B%
}=\{\mathrm{E}_{i}\in \mathcal{A}|i\in I\}$ is a $\sigma $--partition $%
\mathrm{M}=\sum_{i\in I}\mathrm{E}_{i}$ of $\mathrm{M}\subseteq \Lambda $
and $\lambda (\mathrm{E}_{i})\neq 0$. \label{th:a2}
\end{theorem}

\textbf{Proof.} By the classical Ito's formula 
\begin{eqnarray*}
\mathrm{d}(\hat{e}_{g}^{t}\hat{\Phi}_{t}\{X\}) &=&\mathrm{d}\hat{e}_{g}^{t}%
\hat{\Phi}_{t}\{X\}+\hat{e}_{g}^{t}\mathrm{d}\hat{\Phi}_{t}\{X\}+\mathrm{d}%
\hat{e}_{g}^{t}\mathrm{d}\hat{\Phi}_{t}\{X\} \\
&=&\int_{\Lambda }g(t,\mathbf{x})\hat{e}_{g}^{t}\hat{\Phi}_{t}\{X+X\bar{L}%
_{t,\mathbf{x}}+\bar{L}_{t,\mathbf{x}}^{\ast }X\}\mathrm{d}Y_{t}(\mathrm{d}%
\mathbf{x}) \\
&&-\hat{e}_{g}^{t}\hat{\Phi}_{t}\{XK_{t}+K_{t}^{\ast }X-L_{t,\mathbf{x}%
}^{\ast }XL_{t,\mathbf{x}}\}\mathrm{d}t \\
&&+\hat{e}_{g}^{t}\hat{\Phi}_{t}\left\{ \int_{\Lambda }(X\bar{L}_{t,\mathbf{x%
}}+\bar{L}_{t,\mathbf{x}}^{\ast }X)g(t,\mathbf{x})\lambda (\mathrm{d}\mathbf{%
x})\right\} \mathrm{d}t
\end{eqnarray*}%
we obtain from (\ref{eq:ccr2.9}) equation (\ref{eq:ccr2.5}) for $\Phi
_{g}^{t}\{X\}=\langle 0|\hat{e}_{g}^{t}\hat{\Phi}_{t}\{X\}|0\rangle $, if we
take into account that $\langle 0|\mathrm{d}Y_{t}(\mathrm{E})|0\rangle =0$, $%
\forall \mathrm{E}\in \mathcal{B}$ and 
\begin{equation*}
\int_{\Lambda }(X\bar{L}_{t,\mathbf{x}}+\bar{L}_{t,\mathbf{x}}^{\ast }X)g(t,%
\mathbf{x})\lambda (\mathrm{d}\mathbf{x})=\int_{\Lambda }(XL_{t,\mathbf{x}%
}+L_{t,\mathbf{x}}^{\ast }X)g(t,\mathbf{x})\lambda (\mathrm{d}\mathbf{x})
\end{equation*}%
due to $\mathcal{B}$--measurability of $g(t,\cdot )$. Hence equation (\ref%
{eq:ccr2.9}) describes the conditional mean value~(\ref{eq:ccr2.8}) in the
Fock representation $\boldsymbol{w}^{t}\mapsto \boldsymbol{Q}^{t}$ with
respect to the probability measure $\mathrm{d}\mu (\boldsymbol{w}^{t})$
induced on $\mathcal{M}_{t}$ by the vacuum state: 
\begin{equation*}
\int \hat{e}_{g}^{t}(\boldsymbol{w}^{t})\hat{\Phi}\{X\}(\boldsymbol{w}^{t})%
\mathrm{d}\mu (\boldsymbol{w}^{t})=\langle 0|\hat{e}_{g}^{t}\hat{\Phi}%
_{t}\{X\}|0\rangle .\hspace{3cm}\Box
\end{equation*}

\textbf{Remark 2.} In the case of the output momentum process, described in
the Schr\"{o}dinger picture by $Y_{t}(\mathrm{E})=V(t,\mathrm{E})$, $\mathrm{%
E}\in \mathcal{B}$, one can obtain in the same way the posterior equation
for the non-normalized linear stochastic map $\hat{\Phi}_{t}$ in the form 
\begin{eqnarray*}
&&\mathrm{d}\hat{\Phi}_{t}\{X\}+\hat{\Phi}_{t}\{XK_{t}+K_{t}^{\ast
}X-\int_{\Lambda }L_{t,\mathbf{x}}^{\ast }XL_{t,\mathbf{x}}\lambda (\mathrm{d%
}\mathbf{x})\}\mathrm{d}t \\
&=&\frac{1}{i}\ \int_{\Lambda }\hat{\Phi}_{t}\{X\bar{L}_{t,\mathbf{x}}-\bar{L%
}_{t,\mathbf{x}}^{\ast }X\}\mathrm{d}Y_{t}(\mathrm{d}\mathbf{x}).
\end{eqnarray*}%
Then by doubling the space $\Lambda $ and considering the time-continuous
measurement of the commutative family $\boldsymbol{Y}_{t,\mp }=\boldsymbol{Q}%
_{\mp }(t)$ as in section \ref{sec:ccr1}, one can obtain the posterior
equation, corresponding to the complex observation $Z_{t}(\mathrm{E})=W(t,%
\mathrm{E})$, $\mathrm{E}\in \mathcal{B}$ of $\boldsymbol{L}_{t}=\left\{
L_{t,\mathbf{x}}|\mathbf{x\in }\Lambda \right\} $:%
\begin{eqnarray}
&&\mathrm{d}\hat{\Phi}_{t}\{X\}+\hat{\Phi}_{t}\{XK_{t}+K_{t}^{\ast
}X-\int_{\Lambda }L_{t,\mathbf{x}}^{\ast }XL_{t,\mathbf{x}}\lambda (\mathrm{d%
}\mathbf{x})\}\mathrm{d}t  \notag \\
&=&\int_{\Lambda }\hat{\Phi}_{t}\{X\bar{L}_{t,\mathbf{x}}\}\mathrm{d}%
Z_{t}^{\ast }(\mathrm{d}\mathbf{x})+\int_{\Lambda }\hat{\Phi}_{t}\{\bar{L}%
_{t,\mathbf{x}}^{\ast }X\}\mathrm{d}Z_{t}(\mathrm{d}\mathbf{x}).
\label{eq:ccr2.10}
\end{eqnarray}

In the case $\mathcal{B=A}$ of complete complex observation this equation
has a factorizable solution $\hat{\Phi}_{t}\{X\}=\hat{F}_{t}^{\ast }X\hat{F}%
_{t}$, $\forall X\in \mathcal{L}$, where $\hat{F}_{t}$, satisfies the
stochastic equation (\ref{eq:ccr2.3}) in the complexified version 
\begin{equation*}
\mathrm{d}\hat{F}_{t}+K_{t}\hat{F}_{t}\mathrm{d}t=\int_{\Lambda }L_{t,%
\mathbf{x}}\hat{F}_{t}\mathrm{d}Z_{t}^{\ast }(\mathrm{d}\mathbf{x}),\;%
\boldsymbol{Z}_{t}=\frac{1}{\sqrt{2}}\ (\boldsymbol{Y}_{t,+}+i\boldsymbol{Y}%
_{t,-}).
\end{equation*}

\section{A continual observation of CCR quasifree diffusion}

Let $\Xi $ be a symplectic $\sharp $-space, i.e.\ a complex space with
involution 
\begin{equation*}
\xi \in \Xi \mapsto \xi ^{\sharp }\in \Xi \ ,\xi ^{\sharp \sharp }=\xi \
,\;\;\;(\sum \lambda _{i}\xi _{i})^{\sharp }=\sum \lambda _{i}^{\ast }\xi
_{i}^{\sharp }\ ,\forall \lambda _{i}\in \mathbb{C}
\end{equation*}%
and skew-symmetric bilinear $\sharp $-form $s:\Xi \times \Xi \rightarrow 
\mathbb{C}$, such that $s(\xi ^{\sharp },\xi )$ is purely imaginary for all $%
\xi \in \Xi $: 
\begin{equation*}
s(\xi ,\xi ^{\sharp })=-s(\xi ^{\sharp },\xi )\ ,s(\xi ^{\sharp }\,\xi
)^{\ast }=s(\xi ,\xi ^{\sharp }).
\end{equation*}%
We denote by $\Re \Xi $ the real space of the vectors $\xi =\xi ^{\sharp }$,
by $\Theta $ a separating space of complex--valued linear functionals $%
\vartheta :\xi \mapsto \vartheta (\xi )$, on $\Re \Xi $ enquiped with the
weak* topology, $\Im \Theta =\{\vartheta \in \Theta \mid \vartheta
+\vartheta ^{\sharp }=0\}$, where $\vartheta ^{\sharp }(\xi )=\vartheta (\xi
)^{\ast }$, $\forall \xi \in \Re \Xi $ and by $R(\xi ),\xi \in \Xi $ an
operator $\sharp $-representation $R(\xi )^{\ast }=R(\xi ^{\sharp })$\ of
the canonical commutation relations (CCR) 
\begin{equation}
\lbrack R(\xi ),R(\xi ^{\sharp })]=\frac{1}{i}s(\xi ,\xi ^{\sharp }),\
\forall \xi \in \Xi  \label{eq:ccrconcom}
\end{equation}%
in a Hilbert space $\mathcal{H}$\thinspace\ associated with a Gaussian state 
\begin{equation}
\phi _{0}\{e^{iR(\xi )}\}=e^{i\vartheta _{0}(\xi )-\frac{1}{2}\xi
^{2}}\equiv \phi _{0}(\xi ).  \label{eq:ccrgaussian}
\end{equation}%
Here $i\vartheta _{0}\in \Im \Theta $ is defined by the expectation $%
\vartheta _{0}(\xi )=\phi _{0}\{R(\xi )\}$ of $R$ and the quadratic form $%
\xi ^{2}=\langle \xi ,\xi \rangle ,$\thinspace\ satisfying the Heisenberg
inequality 
\begin{equation*}
\xi ^{2}\eta ^{2}-\langle \xi ,\eta \rangle ^{2}\geq \frac{1}{4}\ s(\xi
,\eta )^{2},\hspace{7mm}\forall \xi ,\eta \in \Re \Xi ,
\end{equation*}%
is defined by the symmetric covariance form 
\begin{equation*}
\langle \xi ,\eta \rangle =\frac{1}{2}\ \phi _{0}\{R(\xi )R(\eta )+R(\eta
)R(\xi )\}-\vartheta _{0}(\xi )\vartheta _{0}(\eta ).
\end{equation*}

One can realise $R(\xi)-\vartheta_{0}(\xi)$\ as double real part $2\Re
A_{0}=(A_{0}+A_{0}^{\ast})(\xi)$\ of the creation operator $%
A_{0}^{\ast}(\xi)=A_{0}(\xi^{\sharp})^{\ast}$\ with the vacuum state $%
\phi_{0}\{X\}=(\psi_{0}| X\psi_{0})$\ in an initial Fock space $\mathcal{H=F}%
_{0}$\ over the Hilbert space $\mathrm{H}=\Xi^{\ast}$ , associated with the
scalar product 
\begin{equation*}
(\xi|\eta)=\langle\eta,\xi^{\sharp}\rangle+\frac{i}{2}\ s(\eta,\xi^{\sharp}),%
\hspace{7mm}\forall\xi,\eta\in\Xi.
\end{equation*}
Indeed, the adjoint operators $A_{0}(\xi^{\sharp})\,,A_{0}^{\ast}(\xi)$
satisfying the CCR 
\begin{equation*}
[A_{0}(\xi^{\sharp}),A_{0}^{\ast}(\xi)]=(\xi|\xi),
\end{equation*}
generate $\mathcal{F}_{0}$ by the unitary representation 
\begin{equation}
X(\xi)=e^{i\vartheta_{0}(\xi)-\xi^{2}/2}e^{iA_{0}^{\ast}
(\xi)}e^{iA_{0}(\xi)}\;,\xi\in\Re\Xi  \label{eq:ccrunitary}
\end{equation}
of the Weyl operators $X(\xi)=\exp\{iR(\xi)\}$\ on $\psi_{0}$:

\begin{eqnarray*}
X(\xi )\psi _{0} &=&\phi _{0}(\xi )e^{iA_{0}^{\ast }(\xi )}\psi _{0}\;, \\
X(\eta )X(\xi ) &=&e^{is(\eta ,\xi )}X(\eta +\xi )\;,
\end{eqnarray*}%
and $(\psi _{0}|X(\xi )\psi _{0})=\phi _{0}(\xi )$.

We shall identify the dual space $\Theta$\ with the completion of $\Xi$ in
the (nondegenerate) norm $\mathsf{I}\xi\mathsf{I}=\langle\xi^{\sharp},\xi%
\rangle^{1/2}=\sqrt{(\Re\xi)^{2}+(\Im\xi)^{2}}$ on $\Xi$, such that $%
\vartheta(\xi)=\langle\xi,\vartheta\rangle$. Denoting $\mathbf{j}:\xi\mapsto%
\mathbf{j}\xi=\xi$ the canonical bounded map $\Xi\rightarrow\mathrm{H}$, 
\begin{eqnarray*}
\|\mathbf{j}\xi\|^{2}=(\xi|\xi)=\mathsf{I}\xi\mathsf{I}^{2}+\frac{i}{2}\
s(\xi,\xi^{\sharp})=\mathsf{I}\xi\mathsf{I}^{2}+s(\Re\xi,\Im\xi) \\
\leq\mathsf{I}\xi\mathsf{I}^{2}+|s(\Re\xi,\Im\xi)|\leq\mathsf{I}\xi\mathsf{I}%
^{2}+2\mathsf{I}\Re\xi\mathsf{I} \mathsf{I}\Im\xi\mathsf{I}\leq 2\mathsf{I}%
\xi\mathsf{I}^{2},
\end{eqnarray*}
one can express the scalar product $(\xi|\eta)$ through the complex metric
bounded operator $\mathbf{g}=\mathbf{j}^{\ast}\mathbf{j}$\ as $%
(\xi|\eta)=\langle\xi^{\sharp},\mathbf{g}\eta\rangle$. Here $\vartheta=%
\mathbf{g}\eta\in\Theta$, $\forall\eta\in\Xi$ is the complex functional $%
\vartheta(\xi)=(\xi^{\sharp}|\eta)= \langle\xi,\mathbf{g}\eta\rangle$
defining together with $\vartheta^{\sharp}(\xi)=\langle\xi^{\sharp},\mathbf{g%
}\eta\rangle^{\ast}= (\eta|\xi)$ the $\sharp$-functional $%
2\Re\vartheta=\vartheta+\vartheta ^{\sharp}=2\Re\eta+\mathbf{s}\Im\eta$,
where $\mathbf{s}:\Re\Xi\rightarrow\Theta$ is the skew-symmetric operator $%
\langle\xi,\mathbf{s}\eta\rangle=s(\eta,\xi)$, $\mathsf{I}\mathbf{s}\eta%
\mathsf{I}\leq 2\mathsf{I}\eta\mathsf{I}$ due to the Heisenberg inequality.

Let us consider the quantum diffusion of CCR algebra under the continuous
measurement of the unbounded operators $L_{x}=R(\zeta _{x}),\;x\in \mathbb{R}%
_{+}\times \Lambda $, defined by a family $\{\zeta _{x}|x\in \mathbb{R}%
_{+}\times \Lambda \}$ of vectors in $\Xi $, weakly square integrable: $%
\int_{0}^{t}\varepsilon _{\tau }(\vartheta ^{\sharp },\vartheta )\mathrm{d}%
\tau <\infty $\ for all $t\in \mathbb{R}_{+}$ and $\vartheta \in \Theta $,
where 
\begin{equation}
\varepsilon _{t}(\vartheta ^{\sharp },\vartheta )=\int_{\Lambda }\vartheta
^{\sharp }(\zeta _{t,\mathbf{x}})\vartheta (\zeta _{t,\mathbf{x}}^{\sharp
})\lambda (\mathrm{d}\mathbf{x})=\int_{\Lambda }|\langle \zeta _{t,\mathbf{x}%
}^{\sharp },\vartheta \rangle |^{2}\lambda (\mathrm{d}\mathbf{x}),
\label{eq:ccrwhere}
\end{equation}%
$\vartheta ^{\sharp }(\zeta )=\vartheta ^{\sharp }(\Re \zeta )+i\vartheta
^{\sharp }(\Im \zeta )=\vartheta (\zeta ^{\sharp })^{\ast }$ for all $\zeta
\in \Xi $. Moreover, we shall suppouse that the integral~(\ref{eq:ccrwhere})
is a weak* continuous function of $\vartheta \in \Theta $ such that 
\begin{equation*}
\int_{\Lambda }f(\mathbf{x})\vartheta (\zeta _{t,\mathbf{x}}^{\sharp
})\lambda (\mathrm{d}\mathbf{x})=\langle \boldsymbol{\zeta }_{t}^{\sharp }%
\boldsymbol{f},\vartheta \rangle
\end{equation*}%
for every square integrable function $\boldsymbol{f}:\Lambda \rightarrow 
\mathbb{C}$, where $\boldsymbol{\zeta }_{t}^{\sharp }\boldsymbol{f}$ is an
element of $\Xi $ denoted as $\int_{\Lambda }\zeta _{t,\mathbf{x}}^{\sharp
}f(\mathbf{x})\lambda (\mathrm{d}\mathbf{x})$. We shall suppose also that
the Hamiltonian $H_{t}$ of the system under the observation is given in the
Fock space $\mathcal{H}$\ by the normal ordering $H_{t}=:h_{t}(R):$ of a
quadratic form $v_{t}(\vartheta )+\frac{1}{2}\ \omega _{t}(\vartheta
,\vartheta )=h_{t}(\vartheta )$. This means 
\begin{equation*}
(\psi _{\eta }|H_{t}\psi _{\eta })=v_{t}(\vartheta )+\frac{1}{2}\ \omega
_{t}(\vartheta ,\vartheta ),\;\vartheta =\vartheta _{0}+2\Re (\mathbf{g}\eta
),
\end{equation*}%
where $\psi _{\eta }=\exp \{-\frac{1}{2}\ (\eta |\eta )+A_{0}^{\ast }(\eta
)\}\psi _{0},\;\psi _{0}\in \mathcal{H}$ is the normalized vacuum: $%
A_{0}\psi _{0}=0$, $(\psi _{0}\mid \psi _{0})=1$ in the initial space $%
\mathcal{H}=\mathcal{F}_{0}$.

Let us suppose that $\{v_{t}|t\in \mathbb{R}_{+}\}$ is a locally integrable
family of $\sharp $--linear forms $v_{t}(\vartheta )=\langle \vartheta
,v_{t}\rangle $ and $\{\omega _{t}|t\in \mathbb{R}_{+}\}$ is a locally
integrable family of real symmetric bilinear forms on $\Im \Theta $ such
that 
\begin{equation*}
\mathsf{I}\upsilon \mathsf{I}_{1}^{t}=\int_{0}^{t}\mathsf{I}\upsilon _{r}%
\mathsf{I}\mathrm{d}r<\infty ,\;\;\mathsf{I}\omega \mathsf{I}%
_{1}^{t}=\int_{0}^{t}\mathsf{I}\omega _{r}\mathsf{I}\mathrm{d}r<\infty
\;\;\;\forall t<\infty ,
\end{equation*}%
where $\mathsf{I}\upsilon _{t}\mathsf{I}=\sqrt{\upsilon _{t}^{2}}$, $%
\upsilon _{t}^{2}=\langle \upsilon _{t},\upsilon _{t}\rangle $, $\mathsf{I}%
\omega _{t}\mathsf{I}=\mathrm{sup}\{\omega _{t}(\vartheta ^{\prime
},\vartheta )\mid \mathsf{I}\vartheta ^{\prime }\mathsf{I}<1,\mathsf{I}%
\vartheta \mathsf{I}<1\}$. Assuming the weak* continuity of the linear
functions $\upsilon _{t}(\vartheta )$ and $\omega _{t}(\vartheta ^{\prime
},\vartheta )$ on $\Theta \ni \vartheta ,\;\forall \vartheta ^{\prime }\in
\Theta $, we identify $\upsilon _{t}(\vartheta )$ with $\vartheta (\upsilon
_{t})$, $\upsilon _{t}\in \Re \Xi $ and $\omega _{t}(\vartheta ^{\prime
},\vartheta )$, with $\vartheta ^{\prime }(\boldsymbol{\omega }_{t}\vartheta
)=\langle \vartheta ^{\prime },\boldsymbol{\omega }_{t}\vartheta \rangle $,
where $\boldsymbol{\omega }_{t}$ is a symmetric and hence bounded operator
on the Hilbert space $\Theta $. The quadratic form of $H_{t}$, corresponding
to 
\begin{equation*}
i[H_{t},R(\xi )]=\upsilon _{t}(\mathbf{s}\xi )+R(\boldsymbol{\omega }_{t}%
\mathbf{s}\xi )\;,\;\;\forall \xi \in \Xi
\end{equation*}%
gives together with $i[R(\xi ),L_{t,\mathbf{x}}]=s(\xi ,\zeta _{t,\mathbf{x}%
}),\;i[L_{t,x}^{\ast },R({\xi })]=s(\zeta _{t,\mathbf{x}}^{\sharp },\xi )$
the linear quantum Langevin equation (\ref{eq:ccr2.1}) for $X(t)=j(t,R(\xi
)) $: 
\begin{equation}
\mathrm{d}R(t,\xi )+R(t,i\boldsymbol{\kappa }_{t}\mathbf{s}\xi )\mathrm{d}t=%
\mathrm{d}P(t,\xi )+\upsilon _{t}(\mathbf{s}\xi )\mathrm{d}t.
\label{eq:ccrtripiat}
\end{equation}%
Here $\mathrm{d}P(t,\xi )=i\int_{\Lambda }\{s(\xi ,\zeta _{t,\mathbf{x}%
}^{\sharp })\mathrm{d}A(t,\mathrm{d}\mathbf{x})-s(\xi ,\zeta _{t,\mathbf{x}})%
\mathrm{d}A^{\ast }(t,\mathrm{d}\mathbf{x})\}$, $\boldsymbol{\kappa }%
_{t}:\Im \Theta \rightarrow \Re \Xi $ is the linear imaginary operator $%
\boldsymbol{\kappa }_{t}=\frac{1}{2}\ \boldsymbol{\gamma }_{t}+i\boldsymbol{%
\omega }_{t}$, where $\boldsymbol{\gamma }_{t}=\boldsymbol{\varepsilon }_{t}-%
\boldsymbol{\varepsilon }_{t}^{\sharp }$ is given by the weak* continuous
function $\gamma _{t}(\vartheta ^{\prime },\vartheta )=2i\Im \varepsilon
_{t}(\vartheta ^{\prime },\vartheta )$ of $\vartheta ,\vartheta ^{\prime
}\in \Im \Theta $, $\vartheta ^{\prime }(\boldsymbol{\kappa }_{t}\vartheta
)=\kappa _{t}(\vartheta ^{\prime },\vartheta )=\langle \vartheta ^{\prime },%
\boldsymbol{\kappa }_{t}\vartheta \rangle $, 
\begin{equation*}
\kappa _{t}(\vartheta ^{\prime },\vartheta )=i\Im \varepsilon _{t}(\vartheta
^{\prime },\vartheta )+i\omega _{t}(\vartheta ^{\prime },\vartheta
),\;\forall \vartheta ,\vartheta ^{\prime }\in \Im \Theta \;.
\end{equation*}

The following theorem gives the solution of the operator equation~(\ref%
{eq:ccrtripiat}) together with an integral of a $\mathcal{B}(\mathbb{R}%
_{+})\otimes \mathcal{B}$-measurable locally squareintegrable function $%
g:\Gamma _{1}\rightarrow \mathbb{R}$ over the differential 
\begin{equation}
\mathrm{d}Y(t,\mathrm{E})=R(t,\zeta _{t}(\mathrm{E})+\zeta _{t}^{\sharp }(%
\mathrm{E}))+\mathrm{d}Q(\mathrm{E}).  \label{eq:ccrtrishest}
\end{equation}%
Here $\zeta _{t}(\mathrm{E})\in \Xi \;,\;\zeta _{t}^{\sharp }(\mathrm{E}%
)=\int_{\mathrm{E}}\zeta _{t,\mathbf{x}}^{\sharp }\lambda (\mathrm{d}\mathbf{%
x})=\zeta _{t}(\mathrm{E})^{\sharp }$ is defined for any $\mathrm{E}$ for
which $\lambda (\mathrm{E})<\infty $ due to weak* continuity on $\Theta \ni
\vartheta $ of the integral $\int_{\mathrm{E}}\vartheta (\zeta _{t,\mathbf{x}%
})\lambda (\mathrm{d}\mathbf{x})$. Note, that the corresponding unitary
quantum stochastic evolution (1.5) with unbounded operator 
\begin{equation*}
K_{t}=\frac{1}{2}\ \int_{\Lambda }R(\zeta _{t,\mathbf{x}}^{\sharp })R(\zeta
_{t,\mathbf{x}})\lambda (\mathrm{d}\mathbf{x})+iH_{t}
\end{equation*}%
exists only if $\langle \psi _{0}|K_{t}|\psi _{0}\rangle =\frac{1}{2}\
\int_{\Lambda }\Vert \zeta _{t,\mathbf{x}}\Vert ^{2}\lambda (\mathrm{d}%
\mathbf{x})\equiv k_{t}(0)<\infty $ for almost all $t$. The Wick symbol $%
k_{t}(\vartheta )=(\psi _{\eta }|K_{t}\psi _{\eta })$ is defined in this
case as 
\begin{equation*}
k_{t}(\vartheta )=k_{t}(0)+i\upsilon _{t}(\vartheta )+\frac{1}{2}\ \left(
\varepsilon _{t}(\vartheta ,\vartheta )+i\omega _{t}(\vartheta ,\vartheta
)\right) ,\;\forall \vartheta =\vartheta _{0}+2\Re (\mathbf{g}\eta ).
\end{equation*}

In this theorem we use the notations $\mathsf{I}\cdot \mathsf{I}_{1}^{t}$, $%
\mathsf{I}\cdot \mathsf{I}_{2}^{t}$ for the norms 
\begin{equation*}
\mathsf{I}\boldsymbol{\kappa }\mathsf{I}_{1}^{t}=\int_{0}^{t}\mathsf{I}%
\boldsymbol{\kappa }_{r}\mathsf{I}\mathrm{d}r,\;\;\mathsf{I}\xi \mathsf{I}%
_{2}^{t}=\left( \int_{0}^{t}\int_{\Lambda }\mathsf{I}\xi _{r,\mathbf{x}}%
\mathsf{I}^{2}\lambda (\mathrm{d}\mathbf{x})\mathrm{d}r\right) ^{1/2},
\end{equation*}%
where $\mathsf{I}\xi \mathsf{I}=\langle \xi ,\xi \rangle ^{1/2}$ for a $\xi
\in \Re \Xi $, and $\mathsf{I}\boldsymbol{\kappa }_{t}\mathsf{I}=\mathsf{I}i%
\boldsymbol{\kappa }_{t}\mathsf{I}$ means the norm $\mathsf{I}\boldsymbol{%
\kappa }_{t}\mathsf{I}=\mathrm{sup}\{\mathsf{I}\kappa _{t}\vartheta \mathsf{I%
}/\mathsf{I}\vartheta \mathsf{I}\}$ of the real operator $i\boldsymbol{%
\kappa }_{t}$ on the Hilbert space $\Theta $. Let us also denote $%
g_{t}(r)=g(r)$, $r<t$, $g_{t}(r)=0$, $f_{t}(r,\mathbf{x})=0=f_{t}^{\ast }(r,%
\mathbf{x}),\;\forall r\geq t$, and 
\begin{equation*}
f_{t}^{\ast }(r,\mathbf{x},\xi )=g(r,\mathbf{x})+is(\xi _{r},\zeta _{r,%
\mathbf{x}}^{\sharp })=f(r,\mathbf{x},\xi ^{\sharp })^{\ast },\;r<t.
\end{equation*}

\begin{theorem}
Let the equations~(\ref{eq:ccrtripiat},\ref{eq:ccrtrishest}) for the quantum
diffusion on the CCR algebra be defined by $\upsilon _{t}\in \Re \Xi $, $%
\boldsymbol{\omega }_{t}:\Theta \rightarrow \Re \Xi $ and $\zeta _{x}\in
\Theta $ such that 
\begin{equation*}
\mathsf{I}\upsilon \mathsf{I}_{1}^{t}<\infty ,\;\mathsf{I}i\boldsymbol{%
\kappa }\mathsf{I}_{1}^{t}<\infty ,\;\mathsf{I}\bar{\zeta}+\bar{\zeta}%
^{\sharp }\mathsf{I}_{2}^{t}<\infty ,\;\forall t\in \mathbb{R}_{+}
\end{equation*}%
where $x\mapsto \bar{\zeta}_{x}\in \Xi $ is a weakly $\mathcal{B}(\mathbb{R}%
_{+})\otimes \mathcal{B}$-measurable function of $x=(t,\mathbf{x})$, defined
by $\bar{\boldsymbol{\zeta }}_{t}\boldsymbol{g}=\boldsymbol{\zeta }_{t}%
\boldsymbol{g}$ for every $\boldsymbol{g}\in L_{\mathcal{B}}^{2}(\Lambda )$.
Then the equation~(\ref{eq:ccrtripiat}) has a unique solution, defined in
the Hilbert space $\mathcal{H}\otimes \mathcal{F=F}_{0}\otimes L^{2}(\Gamma
) $ by the quantum stochastic integral 
\begin{equation}
R(t,\xi )+y(g_{t})=\int_{0}^{t}s(\upsilon _{r},\xi _{r})\mathrm{d}r+R(\xi
(t))+a(f_{t}^{\ast })+a^{\ast }(f_{t})  \label{eq:ccrtrisem}
\end{equation}%
along the trajectories $\xi _{r}=\varphi _{r}^{(g)}(t,\xi ),\;r\in \lbrack
0,t)$ of the backward predual differential equation 
\begin{eqnarray}
-\dot{\xi}_{r}+i\boldsymbol{\kappa }_{r}\mathbf{s}\xi _{r} &=&\int_{\Lambda
}g(r,\mathbf{x})(\bar{\zeta}_{r,\mathbf{x}}+\bar{\zeta}_{r,\mathbf{x}%
}^{\sharp })\lambda (\mathrm{d}\mathbf{x}),\;  \label{eq:ccrtrivosem} \\
\xi _{t} &=&\xi \in \Xi ,\;\xi (t)=\varphi _{0}^{(g)}(t,\xi )=\xi _{0},
\end{eqnarray}%
with $g=0$ corresponding to $y(g_{t})=0$. The output integral 
\begin{equation*}
y(g_{t})=\int_{0}^{t}\int_{\Lambda }g(t,\mathbf{x})\mathrm{d}Y(r,\mathrm{d}%
\mathbf{x})
\end{equation*}%
of a $\mathcal{B}(\mathbb{R}_{+})\times \mathcal{B}$-measurable function $%
g\in L_{\mathcal{B}}^{2}(\Gamma _{1})$ over the differntial (\ref%
{eq:ccrtrishest}) is given also by the quantum stochastic integral (\ref%
{eq:ccrtrisem}) along the trajectory $\xi _{r}=\boldsymbol{\varphi }%
_{r}^{(g)}(t,0)$ of the equation (\ref{eq:ccrtrivosem}) with $\xi =0$,
corresponding to $R(t,\xi )=0$. \label{th:perv}
\end{theorem}

\textbf{Proof}. First we write the weak solution of the equation~(\ref%
{eq:ccrtrivosem}) in the standard form 
\begin{equation*}
\xi _{r}=\boldsymbol{\varphi }_{r}(t)\xi +\int_{r}^{t}\int_{\Lambda }g(s,%
\mathbf{x})\boldsymbol{\varphi }_{r}(s)(\bar{\zeta}_{s,\mathbf{x}}+\bar{\zeta%
}_{s,\mathbf{x}}^{\sharp })\mathrm{d}s\lambda (\mathrm{d}\mathbf{x}),
\end{equation*}%
where $\boldsymbol{\varphi }_{r}(t)\xi =\varphi _{r}^{(0)}(t,\xi )$ is the
solution of the equation~(\ref{eq:ccrtrivosem}) with $g=0$. The resolving
operator $\boldsymbol{\varphi }_{r}(t)$ exists as the chronologically
ordered exponential 
\begin{equation*}
\boldsymbol{\varphi }_{r}(t)=\sum_{n=0}^{\infty }\int \ldots \int_{r\leq
t_{1}<\ldots <t_{n}<t}\boldsymbol{\kappa }_{t_{1}}\mathbf{s}\ldots 
\boldsymbol{\kappa }_{t_{n}}\mathbf{s}\mathrm{d}t_{1}\ldots \mathrm{d}t_{n}
\end{equation*}%
due to the estimate $\mathsf{I}\boldsymbol{\varphi }_{r}(t)\mathsf{I}=\sup \{%
\mathsf{I}\boldsymbol{\varphi }_{r}(t)\xi \mathsf{I}\mid \mathsf{I}\xi 
\mathsf{I}<1\}\leq $ 
\begin{equation*}
\leq \sum_{n=0}^{\infty }\mathsf{I}\mathbf{s}\mathsf{I}^{n}\newline
\int \ldots \int_{0\leq t_{1}\ldots t_{n}<t}\mathsf{I}\boldsymbol{\kappa }%
_{t_{1}}\mathsf{I}\ldots \mathsf{I}\boldsymbol{\kappa }_{t_{n}}\mathsf{I}%
\mathrm{d}t_{1}\ldots \mathrm{d}t_{n}\leq \exp \{\mathsf{I}2\boldsymbol{%
\omega }-i\boldsymbol{\gamma }\mathsf{I}_{1}^{t}\}
\end{equation*}%
because $\mathsf{I}\mathbf{s}\mathsf{I}\leq 2$. Hence one can obtain the
existence of $\langle \xi _{r},\vartheta \rangle $ for every $r\in \lbrack
0,t)$ and $\vartheta \in \Theta $ due to the estimate 
\begin{eqnarray*}
|\langle \xi _{r},\vartheta \rangle | &\leq &|\langle \boldsymbol{\varphi }%
_{r}(t)\xi ,\vartheta \rangle |+\int_{r}^{t}\int_{\Lambda }|g(s,\mathbf{x}%
)\langle \boldsymbol{\varphi }_{r}(s)2\Re \bar{\zeta}_{s,\mathbf{x}%
},\vartheta \rangle |\mathrm{d}s\lambda (\mathrm{d}\mathbf{x}) \\
&\leq &\mathsf{I}\xi \mathsf{I}\mathsf{I}\boldsymbol{\varphi }_{r}^{\top
}(t)\vartheta \mathsf{I}+\mathsf{I}g\mathsf{I}_{2}^{t}\mathsf{I}\bar{\zeta}+%
\bar{\zeta}^{\sharp }\mathsf{I}_{2}^{t}(\int_{r}^{t}\mathsf{I}\boldsymbol{%
\varphi }_{r}^{\top }(s)\vartheta \mathsf{I}^{2}\mathrm{d}s)^{1/2} \\
&\leq &\left( \mathsf{I}\xi \mathsf{I}+\mathsf{I}g\mathsf{I}_{2}^{t}\mathsf{I%
}\bar{\zeta}+\bar{\zeta}^{\sharp }\mathsf{I}_{2}^{t}\sqrt{t-r}\right) 
\mathsf{I}\vartheta \mathsf{I}\exp \{\mathsf{I}2\boldsymbol{\omega }-i%
\boldsymbol{\gamma }\mathsf{I}_{1}^{t}\}.
\end{eqnarray*}%
Now we integrate the left hand side in~(\ref{eq:ccrtrisem}), taking into
account~(\ref{eq:ccrtrishest}) and~(\ref{eq:ccrtrivosem}): 
\begin{eqnarray*}
&&R(t,\xi )+\int_{0}^{t}\int_{\Lambda }g(r,\mathbf{x})\left( R(r,2\Re \zeta
_{r}(\mathrm{d}\mathbf{x}))\mathrm{d}r+\mathrm{d}Q(r,\mathrm{d}\mathbf{x}%
)\right) \\
&=&R(t,\xi )+\int_{0}^{t}\left( 2\Re \left\{ \int_{\Lambda }g(r,\mathbf{x})%
\mathrm{d}A(r,\mathrm{d}\mathbf{x})\right\} -R(r,\dot{\xi}_{r}-i\boldsymbol{%
\kappa }_{r}\mathbf{s}\xi _{r})\mathrm{d}r\right) \\
&=&R(0,\xi _{0})+\!\int_{0}^{t}\!(2\Re \left\{ \!\int_{\Lambda }\!g(r,%
\mathbf{x})\mathrm{d}A(r,\mathrm{d}\mathbf{x})\right\} +\mathrm{d}R(r,\xi
_{r})+R(r,i\boldsymbol{\kappa }_{r}\mathbf{s}\xi _{r})\mathrm{d}r) \\
&=&R(\varphi _{0}^{(g)}(t,\xi ))+\!\int_{0}^{t}\!(2\Re \left\{
\!\int_{\Lambda }\!(g(r,\mathbf{x})+is(\xi _{r},\zeta _{r,\mathbf{x}%
}^{\sharp }))\mathrm{d}A(r,\mathrm{d}\mathbf{x})\right\} +\upsilon _{r}(%
\mathbf{s}\xi _{r})\mathrm{d}r),
\end{eqnarray*}%
where $\mathrm{d}R(r,\xi _{r})$ is the quantum stochastic differential $%
\mathrm{d}R(r,\xi )|_{\xi =\xi _{r}}$, satisfying~(\ref{eq:ccrtripiat}) for $%
t=r$. This proves Theorem~(\ref{th:perv}).$\hfill \Box $

Note that the solution $R(t,\xi )$ of the equation~(\ref{eq:ccrtripiat})
given by~(\ref{eq:ccrtrisem}) for $g=0$ preserves the CCR~(\ref{eq:ccrconcom}%
) and satisfies the nondemolition principle 
\begin{equation}
\lbrack R(t,\xi ),Y_{g}(t)]=0\;\;,\;\forall \xi \in \Xi ,\;g\in L_{\mathcal{B%
}}^{2}(\Gamma _{1}),  \label{eq:ccrtrideviat}
\end{equation}%
where $Y_{g}(t)=\int_{0}^{t}\int_{\Lambda }g(r,\mathbf{x})\mathrm{d}Y(r,%
\mathbf{x})$.

It can be proved by using the quantum Ito's formula and 
\begin{eqnarray*}
\lbrack \mathrm{d}R(t,\xi ),\mathrm{d}R(t,\xi ^{\sharp })] &=&[\mathrm{d}%
P(t,\xi ),\mathrm{d}P(t,\xi ^{\sharp })]=\gamma _{t}(\mathbf{s}\xi ,\mathbf{s%
}\xi ^{\sharp })\mathrm{d}t, \\
\lbrack \mathrm{d}R(t,\xi ),\mathrm{d}Y_{g}(t)] &=&[\mathrm{d}P(t\xi ),%
\mathrm{d}Q_{g}(t)]=is(\xi ,(\boldsymbol{\zeta }_{t}+\boldsymbol{\zeta }%
_{t}^{\sharp })\boldsymbol{g}(t))\mathrm{d}t.
\end{eqnarray*}%
Indeed, if $[R(t,\xi ),\;R(t,\xi ^{\sharp })]=\frac{1}{i}\ s(\xi ,\xi
^{\sharp })$, then from~(\ref{eq:ccrtripiat}) it follows 
\begin{equation*}
\lbrack \mathrm{d}R(t,\xi ),R(t,\xi ^{\sharp })]=\kappa _{t}^{\top }(\mathbf{%
s}\xi ,\mathbf{s}\xi ^{\sharp })\mathrm{d}t,\;[R(t,\xi ),\mathrm{d}R(t\xi
^{\sharp })]=-\kappa _{t}(\mathbf{s}\xi ,\mathbf{s}\xi ^{\sharp })\mathrm{d}%
t,
\end{equation*}%
and $\mathrm{d}[R(t,\xi ),R(t,\xi ^{\sharp })]=(\kappa _{t}^{\top }-\kappa
_{t}+\gamma _{t})(\mathbf{s}\xi ,\mathbf{s}\xi ^{\sharp })\mathrm{d}t=0$; 
\begin{eqnarray*}
\mathrm{d}[R(t,\xi ),Y_{g}(t)] &=&[\mathrm{d}R(t,\xi ),Y_{g}(t)]+[R(t,\xi ),%
\mathrm{d}Y_{g}(t)]+[\mathrm{d}R(t,\xi ),\mathrm{d}Y_{g}(t)] \\
&=&[R(t,\xi ),R(t,(\boldsymbol{\zeta }_{t}+\boldsymbol{\zeta }_{t}^{\sharp })%
\boldsymbol{g}(t)]\mathrm{d}t+is(\xi ,\boldsymbol{\zeta }_{t}+\boldsymbol{%
\zeta }_{t}^{\sharp })\boldsymbol{g}(t))\mathrm{d}t \\
&=&0
\end{eqnarray*}%
if $[R(t,\xi ),Y_{g}(t)]=0$ and, hence, $[\mathrm{d}R(t,\xi ),Y_{g}(t)]=0$.

\textbf{Remark 3.} Let $\Xi =\Xi ^{-}\oplus \Xi ^{+}$ be an orthogonal
decomposition of $\Xi $ with respect to the complex scalar product $\langle
\xi ^{\sharp },\eta \rangle $, such that 
\begin{equation*}
(\xi \oplus \eta )^{\sharp }=\eta ^{\sharp }\oplus \xi ^{\sharp },\;\;s(\xi
^{\sharp },\eta )=0=s(\eta ^{\sharp },\xi ),\;\forall \xi \in \Xi
^{-},\;\eta \in \Xi ^{+}.
\end{equation*}%
One can take $\Xi ^{\mp }$ correspondingly as the negative and positive
subspaces of the Hermitian form $2s(\Re \xi ,\,\Im \xi )=is(\xi ,\xi
^{\sharp })$: 
\begin{equation*}
c(\xi ^{\sharp },\xi ):=is(\xi ^{\sharp },\xi )>0,\;\;\forall \xi \in \Xi
^{-},
\end{equation*}%
which is uniquely defined in the case of nondegeneracy: $s(\xi ,\eta )=0$, $%
\forall \eta \in \Xi \Rightarrow \;\xi =0$, but it is not obligatory. If $%
\zeta _{x}\in \Xi ^{-}$ for almost all $x\in \Gamma _{1}$, and $\omega (\xi
^{\sharp },\eta )=0$ for $\xi \in \Xi ^{-},\;\eta \in \Xi ^{+}$, then the
quantum Langevin equation~(\ref{eq:ccrtripiat}) can be written in the
complex linear form 
\begin{equation}
\mathrm{d}L(t,\xi )+L(t,\boldsymbol{\kappa }_{t}\mathbf{c}\xi )\mathrm{d}t=%
\mathrm{d}A(t,\xi )+\eta _{t}(\mathbf{c}\xi )\mathrm{d}t,  \label{eq:ccr3.10}
\end{equation}%
where $\mathrm{d}A(t,\xi )=i\int_{\Lambda }s(\xi ,\zeta _{t,\mathbf{x}%
}^{\sharp })\mathrm{d}A(t,\mathrm{d}\mathbf{x})$, $L(t,\xi )=R(t,\xi )$, $%
\mathbf{c}\xi =i\boldsymbol{s}\xi $, $\eta _{t}(\mathbf{c}\xi )=\upsilon
_{t}(\mathbf{s}\xi ),\;\forall \xi \in \Xi ^{-}$ and $L(t,\xi )=0$, $\mathrm{%
d}A(t,\xi )=0$, $\mathbf{c}\xi =0$, $\eta _{t}(\mathbf{c}\xi )=0$, $\forall
\xi \in \Xi ^{+}$. The equation~(\ref{eq:ccr1.10}) for a complex observation
can be written in these terms as 
\begin{equation*}
\mathrm{d}Z(t,\mathrm{E})=L(\zeta (\mathrm{E}))\mathrm{d}t+\mathrm{d}W(t,%
\mathrm{E}),\;\;\mathrm{E}\in \mathcal{B}.
\end{equation*}

\section{A CCR quasi-free posterior dynamics and continuous collapse}

The solution~(\ref{eq:ccrtrisem}) obtained for the quasi-free diffusion (\ref%
{eq:ccrtripiat}) with the continuous observation (\ref{eq:ccrtrishest}) of
CCR gives the possibility to solve easily the equation (\ref{eq:ccr2.5}) at
least for the initial Weyl operators $\Phi _{g}^{0}\{X\}=e^{iR(\xi )}=X(\xi
) $. To this end let us represent the product $X(\xi )\otimes \hat{e}%
_{g}^{t} $ of the operator (\ref{eq:ccrunitary}) on $\mathcal{H}=\mathcal{F}%
_{0}$ and the Wick exponent $\hat{e}_{g}^{t}=e^{q(g_{t})-g_{t}^{2}/2}$ of
the integral $y_{t}(g)=q(g_{t})$ on $\mathcal{F}=L^{2}(\Gamma )$ as the
exponent of an operator in Heisenberg picture : 
\begin{equation*}
G(t,\xi )=e^{b^{\ast }(g_{t})}X(t,\xi )e^{b(g_{t})}=e^{R(t,i\xi
)+y(g_{t})-g_{t}^{2}/2},
\end{equation*}%
where $y(g_{t})=\int_{0}^{t}\int_{\Lambda }g(r,\mathbf{x})\mathrm{d}Y(r,%
\mathrm{d}\mathbf{x})$. Due to~(\ref{eq:ccrtrisem}) the exponent 
\begin{equation*}
R(t,i\xi )+y(g_{t})=\left\{ \int_{0}^{t}s(\upsilon _{r},\varphi
_{r}^{(g)}(t))\mathrm{d}r+R(\varphi _{0}^{(g)}(t))+a(f_{t}^{\ast })+a^{\ast
}(f_{t})\right\} (i\xi )
\end{equation*}%
can be written in normally ordered form with respect to 
\begin{equation*}
a^{\ast }(f(i\xi ))=a^{\ast }(g-is(\varphi ^{(g)}(t,i\xi ),\zeta )),\
a(f^{\ast }(i\xi ))=a(g-is(\zeta ^{\sharp },\varphi ^{(g)}(t,i\xi )))
\end{equation*}%
as 
\begin{eqnarray*}
G(t,\xi ) &=&c_{g}(t)\exp \{R(\varphi _{0}^{(g)}(t))+a(f_{t}^{\ast
})+a^{\ast }(f_{t})\}(i\xi )= \\
&=&e^{I_{g}^{t}(i\xi )}e^{a^{\ast }(f_{t}(i\xi ))}X(\frac{1}{i}\ \varphi
_{0}^{(g)}(t,i\xi ))e^{a(f_{t}^{\ast }(i\xi ))}\;.
\end{eqnarray*}%
Here $c_{g}(t)=\exp \{\int_{0}^{t}s(\upsilon _{r},\xi _{r}\mathrm{d}%
r-g_{t}^{2}/2\},\;g_{t}^{2}=\int_{0}^{t}\int_{\Lambda }g(r,\mathbf{x})^{2}%
\mathrm{d}r\lambda (\mathrm{d}\mathbf{x})$ and 
\begin{equation*}
I_{g}^{t}(\xi )=\ln c_{g}(t)+\frac{1}{2}\ |f_{t}|^{2}(\xi
),\;|f_{t}|^{2}(\xi )=\int_{0}^{t}\int_{\Lambda }f^{\ast }(r,\mathbf{x},\xi
)f(r,\mathbf{x},\xi )\mathrm{d}r\lambda (\mathrm{d}\mathbf{x})
\end{equation*}%
is given by an integral over the trajectories $\xi _{r}=\varphi
_{r}^{(g)}(t,\xi )$: 
\begin{equation}
I_{g}^{t}(\xi )=\int_{0}^{t}\{s\left( \upsilon _{r}+\Im \bar{\boldsymbol{%
\zeta }}_{r}^{\sharp }\boldsymbol{g}(r),\;\xi _{r}\right) +\frac{1}{2}\
\varepsilon _{r}(\boldsymbol{s}\xi _{r},\boldsymbol{s}\xi _{r})\}\mathrm{d}%
r\;,  \label{eq:ccrtragec}
\end{equation}%
where $\bar{\boldsymbol{\zeta }}_{t}^{\sharp }\boldsymbol{g}%
(t)=\int_{\Lambda }\bar{\zeta}_{t,\mathbf{x}}^{\sharp }g(t,\mathbf{x}%
)\lambda (\mathrm{d}\mathbf{x})=\boldsymbol{\zeta }_{t}^{\sharp }\boldsymbol{%
g}(t)$ for every $\mathcal{B}$-measurable function $\boldsymbol{g}(t):%
\mathbf{x}\in \Lambda \mapsto g(t,\mathbf{x})$. Hence the operator-function $%
\Phi _{g}^{t}(\xi )=\Phi _{g}^{t}\{X(\xi )\}$, being the vacuum expectation $%
\langle 0|G(t,\xi )|0\rangle $ is defined in $\mathcal{H}$ by 
\begin{equation}
\Phi _{g}^{t}(\xi )=\exp \{I_{g}^{t}(i\xi )+R(\varphi _{0}^{(g)}(t,i\xi ))\},
\label{eq:ccrhence}
\end{equation}%
since $e^{a(f_{t}^{\ast })}|0\rangle =|0\rangle $ for every $f$, where $%
f_{t}^{\ast }(x)=f^{\ast }(x)$ on $x\in \lbrack 0,t)\times \Lambda $ and $%
f_{t}(r,\mathbf{x})=0$, if $r>t$. One can easily verify that (\ref%
{eq:ccrhence}) satisfies the equation (\ref{eq:ccr2.5}), written in the CCR
quasi-free case for $X=X(\xi )$ in the differential form 
\begin{eqnarray}
&&i\frac{\mathrm{d}}{\mathrm{d}t}\ \!\Phi _{g}^{t}(\xi )+\{s(\upsilon
_{r},\xi )-\langle \boldsymbol{\kappa }_{t}\mathbf{s}\xi ,\partial \rangle +%
\frac{i}{2}\ \varepsilon _{t}(\mathbf{s}\xi ,\mathbf{s}\xi )\}\Phi
_{g}^{t}(\xi )  \notag \\
&=&2\Im \langle \bar{\boldsymbol{\zeta }}_{t}\boldsymbol{g}(t),i\partial +%
\frac{1}{2}\ \mathbf{s}\xi \rangle \Phi _{g}^{t}(\xi ).  \label{eq:ccrdiform}
\end{eqnarray}%
This form of the main equation follows from the relations 
\begin{equation*}
\lbrack R(\zeta ),X(\xi )]=s(\zeta ,\xi )X(\xi ),\;\frac{i}{2}\ (X(\xi
)R(\zeta )+R(\zeta )X(\xi ))=\langle \zeta ,\partial \rangle X(\xi )
\end{equation*}%
defining the derivative $\langle \zeta ,\partial \rangle $ of $X(\xi
)=e^{iR(\xi )}$ and the right hand side in (\ref{eq:ccr2.5}) by 
\begin{equation*}
R(\zeta )^{\ast }X(\xi )+X(\xi )R(\zeta )=\frac{2}{i}\Im \langle \zeta
,i\partial +\frac{1}{2}\ \mathbf{s}\xi \rangle X(\xi ),
\end{equation*}%
and also from the definition of the quasi-free Hamiltonian evolution in
terms of the Weyl operators (\ref{eq:ccrunitary}): 
\begin{equation*}
\lbrack H,X(\xi )]=\{s(\upsilon _{t},\xi )-\langle \boldsymbol{\omega }_{t}%
\mathbf{s}\xi ,i\partial \rangle \}X(\xi )\;.
\end{equation*}%
Thus we obtain the solution $\mathrm{M}^{t}(\xi )=\Phi _{0}^{t}(\xi )$ of
the Lindblad equation for the CCR quasi-free case in term of the
characteristic operator $\mathrm{M}^{t}\{X(\xi )\}$ of a prior dynamical map 
$\phi _{0}\mapsto \phi _{0}\mathrm{M}^{t}$, having the differential form~(%
\ref{eq:ccrdiform}) with zero right hand side , as well as the
operator-valued generating functional $\mathrm{P}_{g}=\Phi _{g}^{\infty }(0)$
for the factorial moments of the observable process~(\ref{eq:ccrtrishest})

A posterior quasi-free dynamics of the CCR-algebra under the continual
observation~(\ref{eq:ccrtrishest}) is described by the characteristic a
posteriori function $\hat{\Phi}_{t}(\xi )=\hat{\Phi}_{t}\{X(\xi )\}$ with
the Wick symbol $\langle f|\hat{\Phi}_{t}(\xi )|f\rangle =\Phi _{g}^{t}(\xi
),\;g=\bar{f}+\bar{f}^{\ast }$ of the Gaussian form~(\ref{eq:ccrhence}).
Hence the operator-valued function $\hat{\Phi}_{t}(\xi )$, normalised on the
probability density operator $\hat{\mathrm{P}}_{t}=\hat{\Phi}_{t}(0)$, in
the CCR quasi-free case can be represented as the normal ordered functional~(%
\ref{eq:ccrstocheq}) of $y_{t}=b_{t}+b_{t}^{\ast }$ instead of $g=g_{t}$,
where $\hat{\xi}_{r}=\xi _{r}(y)$, instead of $\xi _{r}=\varphi
_{r}^{(g)}(t,\xi )$, defined by the solution $\xi _{r}(y)=\varphi
_{r}^{(y)}(t,\xi )$ of the backward stochastic equation 
\begin{equation}
-\mathrm{d}_{-}\hat{\xi}_{r}+i\boldsymbol{\kappa }_{r}s\hat{\xi}_{r}\mathrm{d%
}r=\int_{\Lambda }(\bar{\zeta}_{r,\mathbf{x}}+\bar{\zeta}_{r,\mathbf{x}%
}^{\sharp })\mathrm{d}Y_{r}(\mathrm{d}\mathbf{x}),\;\hat{\xi}_{t}=\xi .
\label{eq:ccrstocheq}
\end{equation}

In order to find the operator $\hat{\Phi}_{t}(\xi )=\Phi _{t}(\xi ,y_{t})$
in the form of a function $\Phi _{t}(\xi ,\upsilon _{t})$ of the
trajectories $\upsilon _{t}(g)=\omega (g_{t})$ of the observable process $%
y_{t}(g)=q(g_{t})$, let us solve the equation (\ref{eq:ccr2.9}) with $%
X=X(\xi )$, having in the quasi-free case the Wick symbol~(\ref{eq:ccrdiform}%
) in $\mathcal{H}=\mathcal{F}_{0}$. It can be done in terms of 
\begin{equation*}
\hat{\phi}_{t}^{\vartheta }(\xi )=(\psi _{\eta }|\hat{\Phi}_{t}(\xi )\psi
_{\eta }),\;\;\vartheta =\vartheta _{0}+2\Re (\mathbf{g}\eta )\;,
\end{equation*}%
by solving the linear stochastic differential equation, coresponding to~(\ref%
{eq:ccrdiform}) 
\begin{eqnarray}
&&i\mathrm{d}\hat{\phi}_{t}(\xi )+\{s(\upsilon _{t},\xi )-\langle 
\boldsymbol{\kappa }_{t}\mathbf{s}\xi ,\partial \rangle +\frac{i}{2}\
\varepsilon _{t}(\mathbf{s}\xi ,\mathbf{s}\xi )\}\hat{\phi}_{t}(\xi )\mathrm{%
d}t  \notag \\
&=&\int_{\Lambda }2\Im \langle \bar{\zeta}_{t,\mathbf{x}},i\partial +\frac{%
\mathbf{s}}{2}\ \xi \rangle \hat{\phi}_{t}(\xi )\mathrm{d}Y_{t}(\mathrm{d}%
\mathbf{x})\;,  \label{eq:ccrposterchar}
\end{eqnarray}%
as the equation for a posterior characteristic functuon $\hat{\phi}_{t}(\xi
)=\phi ^{\vartheta }\{\hat{\Phi}_{t}(\xi )\}\equiv \hat{\phi}_{t}^{\vartheta
}(\xi )$ with a Gaussian $\hat{\phi}_{0}(\xi )=\exp \{i\vartheta (\xi )-\xi
^{2}/2\}\equiv \phi ^{\vartheta }(\xi )$. The stochastic function $\hat{\phi}%
_{t}^{\vartheta }(\xi )$ defines the operator-valued function $\hat{\Phi}%
_{t}(\xi )$ as the normal ordered form :$\hat{\phi}_{t}^{R}(\xi )$: of the
initial operators $R-\vartheta _{0}=A_{0}+A_{0}^{\ast }$ in $\mathcal{H}=%
\mathcal{F}_{0}$.

\begin{theorem}
Let the initial state $\phi _{0}$ of the CCR~(\ref{eq:ccrconcom}) with a
linear quantum stochastic evolution~(\ref{eq:ccrtripiat}) have the Gaussian
characteristic function~(\ref{eq:ccrgaussian}). Then a posteriori
nonnormalised state $\hat{\phi}_{t}(\xi )=\phi _{0}\{\hat{\Phi}_{t}(\xi )\}=%
\hat{\phi}_{t}\{X(\xi )\}$ under the continuous nondemolition observation~(%
\ref{eq:ccrtrishest}) also has a Gaussian form 
\begin{equation}
\hat{\phi}_{t}(\xi )=\hat{\rho}_{t}\exp \{i\hat{\vartheta}_{t}(\xi )-\frac{1%
}{2}\ p_{t}(\xi ,\xi )\}\;.  \label{eq:ccrgausform}
\end{equation}%
Here 
\begin{equation*}
\hat{\rho}_{t}=\exp \int_{0}^{t}\int_{\Lambda }\{\hat{\vartheta}_{r}(2\Re 
\bar{\zeta}_{r,\mathbf{x}})\mathrm{d}Y_{r}(\mathrm{d}\mathbf{x})-\frac{1}{2}%
\ \hat{\vartheta}_{r}(2\Re \bar{\zeta}_{r,\mathbf{x}})^{2}\mathrm{d}r\lambda
(\mathrm{d}\mathbf{x})\}
\end{equation*}%
is the probability density $\hat{\rho}_{t}=\hat{\phi}_{t}(0)=\rho (y_{t})$
of the observation up to time $t,\;\;\hat{\vartheta}_{t}(\xi )=\langle \xi ,%
\hat{\vartheta}_{t}\rangle $ is the linear stochastic functional $\hat{%
\vartheta}_{t}=\vartheta _{t}(y_{t})$ of the posterior mean value of $%
R(t,\xi )$, satisfying the linear filtering equation 
\begin{equation}
\mathrm{d}\hat{\vartheta}_{t}(\xi )+\hat{\vartheta}_{t}(i\boldsymbol{\kappa }%
_{t}\mathbf{s}\xi )\mathrm{d}t=\int_{\Lambda }2\Re \langle \xi ,\mathbf{k}%
_{t}\bar{\zeta}_{t,\mathbf{x}}^{\sharp }\rangle \mathrm{d}\tilde{Y}_{t}(%
\mathrm{d}\mathbf{x})+\upsilon _{t}(\mathbf{s}\xi )\mathrm{d}t
\label{eq:ccrfilter}
\end{equation}%
with $\hat{\vartheta}_{0}=\vartheta _{0},\;\mathrm{d}\tilde{Y}_{t}(\mathrm{d}%
x)=\mathrm{d}Y_{t}(\mathrm{d}x)-\mathrm{d}t\int_{\Lambda }\hat{\vartheta}%
_{t}\left( 2\Re \bar{\zeta}_{t,\mathbf{x}}\right) \lambda (\mathrm{d}\mathbf{%
x}),\;\mathbf{k}_{t}=\mathbf{p}_{t}+\!\frac{i}{2}\mathbf{s}$, and $p_{t}(\xi
,\xi )=\langle \xi ,\mathbf{p}_{t}\xi \rangle $ is the quadratic form of the
posterior covariance of $R(t,\xi )$, satisfying the Riccati equation with $%
p_{0}(\xi ,\xi )=\langle \xi ,\xi \rangle $: 
\begin{equation}
\frac{\mathrm{d}}{\mathrm{d}t}\ \!p_{t}(\xi ,\xi )+2p_{t}(\xi ,i\boldsymbol{%
\kappa }_{t}\mathbf{s}\xi )=\varepsilon _{t}(\mathbf{s}\xi ,\mathbf{s}\xi
)-\int_{\Lambda }|2\Re \langle \xi ,\mathbf{k}_{t}\bar{\zeta}_{t,\mathbf{x}%
}^{\sharp }\rangle |^{2}\lambda (\mathrm{d}\mathbf{x}).
\label{eq:ccrrikkati}
\end{equation}%
\label{th:third}
\end{theorem}

\textbf{Proof.} Let us find from~(\ref{eq:ccrposterchar}) a stochastic
equation for $\hat{\lambda}_{t}(i\xi )=\ln \hat{\phi}_{t}(\xi )$, using the
Ito's formula $\mathrm{d}\hat{\lambda}_{t}(i\xi )=\hat{\phi}_{t}^{-1}\mathrm{%
d}\hat{\phi}_{t}-\frac{1}{2}\ (\mathrm{d}\hat{\lambda}_{t}(i\xi ))^{2}$, and 
\begin{eqnarray*}
(\mathrm{d}\hat{\lambda}_{t}(i\xi ))^{2} &=&(\hat{\phi}_{t}^{-1}\mathrm{d}%
\hat{\phi}_{t})^{2}=\mathrm{d}t\int_{\Lambda }\{\langle 2\Re \bar{\zeta}_{t,%
\mathbf{x}},\hat{\lambda}_{t}^{\prime }(i\xi )\rangle +\langle i\Im \bar{%
\zeta}_{t,\mathbf{x}}^{\sharp },\mathbf{s}\xi \rangle \}^{2}\lambda (\mathrm{%
d}\mathbf{x}) \\
&=&\{\bar{\mu}_{t}(\hat{\lambda}_{t}^{\prime }(i\xi ),\hat{\lambda}%
_{t}^{\prime }(i\xi ))+2\bar{\kappa}_{t}(\hat{\lambda}_{t}^{\prime }(i\xi ),%
\mathbf{s}\xi )-\bar{\nu}(\mathbf{s}\xi ,\mathbf{s}\xi )\}\mathrm{d}t
\end{eqnarray*}%
due to $(\mathrm{d}Y_{t}(\mathrm{d}\mathbf{x}))^{2}=\mathrm{d}t\lambda (%
\mathrm{d}\mathbf{x})$, where $\hat{\lambda}_{t}^{\prime }(\xi )=\partial 
\hat{\lambda}_{t}(\xi )$, 
\begin{eqnarray*}
\bar{\mu}_{t}(\vartheta ,\vartheta ) &=&\int_{\Lambda }\langle 2\Re \bar{%
\zeta}_{t,x},\vartheta \rangle ^{2}\lambda (\mathrm{d}\mathbf{x}),\;\bar{\nu}%
_{t}(\mathbf{s}\xi ,\mathbf{s}\xi )=\int_{\Lambda }\langle \Im \bar{\zeta}%
_{t,\mathbf{x}},\mathbf{s}\xi \rangle ^{2}\lambda (\mathrm{d}\mathbf{x}), \\
\bar{\kappa}_{t}(\vartheta ,\mathbf{s}\xi ) &=&\int_{\Lambda }\langle 2\Re 
\bar{\zeta}_{t,\mathbf{x}},\vartheta \rangle \langle i\Im \bar{\zeta}_{t,%
\mathbf{x}}^{\sharp },\mathbf{s}\xi \rangle \lambda (\mathrm{d}\mathbf{x}%
)=\langle \tilde{\boldsymbol{\kappa }}_{t}\mathbf{s}\xi ,\vartheta \rangle .
\end{eqnarray*}%
It gives a quasilinear stochastic equation of the first order for $\hat{%
\lambda}_{t}(\xi )$: 
\begin{eqnarray*}
&&\mathrm{d}\hat{\lambda}_{t}+\{s(\xi ,\upsilon _{t})+\langle i\tilde{%
\boldsymbol{\kappa }}_{t}\mathbf{s}\xi ,\hat{\lambda}_{t}^{\prime }\rangle -%
\frac{1}{2}\ \tilde{\varepsilon}_{t}(\mathbf{s}\xi ,\mathbf{s}\xi )\}\mathrm{%
d}t \\
&=&\int_{\Lambda }\{\langle 2\Re \bar{\zeta}_{t,\mathbf{x}},\hat{\lambda}%
_{t}^{\prime }\rangle +\langle \Im \bar{\zeta}_{t,\mathbf{x}}^{\sharp },%
\mathbf{s}\xi \rangle \}\mathrm{d}Y_{t}(\mathrm{d}\mathbf{x})-\frac{1}{2}\ 
\bar{\mu}_{t}(\hat{\lambda}_{t}^{\prime },\hat{\lambda}_{t}^{\prime })%
\mathrm{d}t\;,
\end{eqnarray*}%
where $\tilde{\boldsymbol{\kappa }}_{t}=\boldsymbol{\kappa }_{t}-\bar{%
\boldsymbol{\kappa }}_{t}$ and $\tilde{\boldsymbol{\varepsilon }}_{t}=%
\boldsymbol{\varepsilon }_{t}-\bar{\boldsymbol{\varepsilon }}_{t}$. This
equation has a quadratic form solution 
\begin{equation*}
\hat{\lambda}_{t}(\xi )=\ln \hat{\rho}_{t}+\hat{\vartheta}_{t}(\xi )+\frac{1%
}{2}\ p_{t}(\xi ,\xi ),\;\hat{\lambda}_{t}^{\prime }(\xi )=\hat{\vartheta}%
_{t}+\mathbf{p}_{t}\xi ,\;\hat{\lambda}_{t}^{\prime \prime }=\mathbf{p}_{t},
\end{equation*}%
where 
\begin{eqnarray*}
\mathrm{d}\ln \hat{\rho}_{t} &=&\mathrm{d}\hat{\lambda}_{t}(0)=\int_{\Lambda
}\langle 2\Re \bar{\zeta}_{t,\mathbf{x}},\hat{\vartheta}_{t}\rangle \mathrm{d%
}Y_{t}(\mathrm{d}\mathbf{x})-\frac{1}{2}\ \bar{\mu}_{t}(\hat{\vartheta}_{t},%
\hat{\vartheta}_{t})\mathrm{d}t, \\
\mathrm{d}\hat{\vartheta}_{t}\! &=&\!\mathrm{d}\hat{\lambda}_{t}^{\prime
}(0)\!=\!\int_{\Lambda }\!(2\mathbf{p}_{t}\Re \bar{\zeta}_{t,\mathbf{x}}+%
\mathbf{s}\Im \bar{\zeta}_{t,\mathbf{x}})\mathrm{d}Y_{t}(\mathrm{d}\mathbf{x}%
)-\{(\mathbf{p}_{t}\bar{\mathbf{\mu }}_{t}-\!i\mathbf{s}\boldsymbol{\kappa }%
_{t}^{\top })\hat{\vartheta}_{t}-\!\mathbf{s}\upsilon _{t}\}\mathrm{d}t, \\
\mathrm{d}\mathbf{p}_{t} &=&\mathrm{d}\hat{\lambda}_{t}^{\prime \prime }=-\{%
\mathbf{p}_{t}\bar{\boldsymbol{\mu }}_{t}\mathbf{p}_{t}+\mathbf{s}\tilde{%
\boldsymbol{\varepsilon }}_{t}\mathbf{s}+i(\mathbf{p}_{t}\tilde{\boldsymbol{%
\kappa }}_{t}\mathbf{s}-\mathbf{s}\tilde{\boldsymbol{\kappa }}_{t}^{\top }%
\mathbf{p}_{t})\}\mathrm{d}t\;,
\end{eqnarray*}%
where $\tilde{\kappa}^{\top }(\mathbf{s}\xi ,\mathbf{p}\xi )=\tilde{\kappa}(%
\mathbf{p}\xi ,\mathbf{s}\xi )$. Using the integral form of the symmetric $%
\ast $-weakly continuous operators $\bar{\boldsymbol{\mu }}_{t},\;\bar{%
\boldsymbol{\nu }}_{t}:\Theta \rightarrow \Xi $, one can obtain the
stochastic integral representation of $\ln \hat{\rho}_{t}=\int_{0}^{t}%
\mathrm{d}\hat{\lambda}_{t}(0)$ in Theorem~\ref{th:third} as well as the
equation~(\ref{eq:ccrfilter}) and~(\ref{eq:ccrrikkati}) for $\hat{\vartheta}%
_{t}(\xi )=\langle \xi ,\hat{\vartheta}_{t}\rangle $ and $p_{t}(\xi ,\xi
)=\langle \xi ,\mathbf{p}_{t}\xi \rangle $ with $\mathbf{k}_{t}=\mathbf{p}%
_{t}+\frac{i}{2}\ \!\mathbf{s}$ due to $\mathbf{s}^{\top }=-\mathbf{s}$, 
\begin{eqnarray*}
\bar{\mu}_{t}(\mathbf{p}_{t}\xi ,\mathbf{p}_{t}\xi )-2i\bar{\kappa}_{t}(%
\mathbf{p}_{t}\xi ,\mathbf{s}\xi )+\bar{\nu}_{t}(\mathbf{s}\xi ,\mathbf{s}%
\xi ) &=&\!\int_{\Lambda }\!(2\Re \langle \xi ,(\mathbf{p}_{t}+\frac{i}{2}\
\!\mathbf{s})\bar{\zeta}_{t,\mathbf{x}}^{\sharp }\rangle )^{2}\lambda (%
\mathrm{d}\mathbf{x}); \\
2\Re \langle \xi ,(\mathbf{p}_{t}+\frac{i}{2}\ \mathbf{s})\bar{\zeta}_{t,%
\mathbf{x}}^{\sharp }\rangle &=&\langle \xi ,2\mathbf{p}_{t}\Re \bar{\zeta}%
_{t,\mathbf{x}}+\mathbf{s}\Im \bar{\zeta}_{t,\mathbf{x}}\rangle ,\;\xi \in
\Re \Xi .
\end{eqnarray*}

Let's point out that quantum filtering equations~(\ref{eq:ccrfilter}), ~(\ref%
{eq:ccrrikkati}), represented in the short form 
\begin{eqnarray}
\mathrm{d}\hat{\vartheta}_{t}(\xi )+\hat{\vartheta}_{t}(\boldsymbol{\alpha }%
_{t}\xi )\mathrm{d}t &=&\int_{\Lambda }2\Re \langle \xi ,\mathbf{k}_{t}\bar{%
\zeta}_{t,\mathbf{x}}^{\sharp }\rangle \mathrm{d}Y_{t}(\mathrm{d}\mathbf{x}%
)+\upsilon _{t}(\mathbf{s}\xi )\mathrm{d}t,  \notag \\
\frac{\mathrm{d}}{\mathrm{d}t}\ \!p_{t}(\xi ,\xi )+2p_{t}(\xi ,\boldsymbol{%
\alpha }_{t}\xi ) &=&\tilde{\varepsilon}_{t}(\mathbf{s}\xi ,\mathbf{s}\xi )+%
\bar{\mu}_{t}(\mathbf{p}_{t}\xi ,\mathbf{p}_{t}\xi ),\;\;\mathbf{p}_{0}=%
\boldsymbol{1},  \label{eq:ccrshort}
\end{eqnarray}%
where $\boldsymbol{\alpha }_{t}=\bar{\boldsymbol{\mu }}_{t}\mathbf{p}_{t}+i%
\tilde{\boldsymbol{\kappa }}_{t}\mathbf{s}$, give $\hat{\lambda}_{t}(\xi
)=\ln \hat{\phi}_{t}(\frac{1}{i}\ \xi )$ in the form of the integral 
\begin{equation}
\hat{\lambda}_{t}(\xi )=\vartheta _{0}(\hat{\xi}_{0})+\frac{1}{2}\ \hat{\xi}%
_{0}^{2}+\int_{0}^{t}\{\langle \mathbf{s}\hat{\xi},\mathrm{d}\Upsilon
_{r}\rangle +\frac{1}{2}\ \!\left( \tilde{\varepsilon}_{r}(\mathbf{s}\hat{\xi%
},\mathbf{s}\hat{\xi})-\bar{\mu}_{r}(\hat{p}_{r}(\hat{\xi}),\hat{p}_{r}(\hat{%
\xi}))\right) \mathrm{d}r\}  \label{eq:ccrintform}
\end{equation}%
over the stochastic trajectories $\hat{\xi}_{r}=\hat{\varphi}_{r}(t,\xi )$
of the backward linear equation~(\ref{eq:ccrstocheq}) with $\hat{p}_{r}(\hat{%
\xi})=\hat{\vartheta}_{r}+\mathbf{p}_{r}\hat{\xi}_{r},\;\hat{\xi}_{0}=\hat{%
\varphi}_{0}(t,\xi )$ and 
\begin{equation*}
\mathrm{d}\Upsilon _{t}=\upsilon _{t}\mathrm{d}t+\Im \int_{\Lambda }\bar{%
\zeta}_{t,\mathbf{x}}^{\sharp }\mathrm{d}Y_{t}(\mathrm{d}\mathbf{x}).
\end{equation*}%
Indeed, if $\mathrm{d}_{-}\hat{\xi}_{r}=\hat{\xi}_{r}-\hat{\xi}_{r-\mathrm{d}%
r}$ is the backward stochastic differential in~(\ref{eq:ccrstocheq}), then $%
\mathrm{d}\langle \hat{\vartheta}_{r},\hat{\xi}_{r}\rangle =\langle \hat{\xi}%
_{r},\mathrm{d}\hat{\vartheta}_{r}\rangle +\langle \hat{\vartheta}_{r},%
\mathrm{d}_{-}\hat{\xi}_{r}\rangle $ and 
\begin{equation*}
\mathrm{d}\!\left( p_{r}(\hat{\xi}_{r},\hat{\xi}_{r})\right) =2p_{r}(\hat{\xi%
}_{r},\mathrm{d}_{-}\hat{\xi}_{r})+\dot{p}_{r}(\hat{\xi}_{r},\hat{\xi}_{r})%
\mathrm{d}r.
\end{equation*}%
Using the equations~(\ref{eq:ccrshort}) and writting the equation (\ref%
{eq:ccrstocheq}) in the form $\mathrm{d}_{-}\hat{\xi}_{r}=\boldsymbol{\alpha 
}_{r}\hat{\xi}_{r}\mathrm{d}r-2\Re \bar{\boldsymbol{\zeta }}_{r}\mathrm{d}%
\hat{\boldsymbol{Y}}_{r}$ with respect to 
\begin{equation*}
\mathrm{d}\hat{Y}_{r}(\mathrm{d}\mathbf{x})=\mathrm{d}Y_{r}(\mathrm{d}%
\mathbf{x})+\langle \mathbf{p}_{r}\hat{\xi}_{r},2\Re \bar{\zeta}_{r,\mathbf{x%
}}\rangle \mathrm{d}r\lambda (\mathrm{d}\mathbf{x})
\end{equation*}%
one can obtain by integrating by parts of the difference $\hat{\lambda}%
_{t}(\xi )-\ln \hat{\rho}_{t}-\vartheta _{0}(\hat{\xi}_{0})-\frac{1}{2}\ 
\hat{\xi}_{0}^{2}$: 
\begin{eqnarray*}
&&\hat{\vartheta}_{t}(\xi )-\vartheta _{0}(\hat{\xi}_{0})+\frac{1}{2}\
p_{t}(\xi ,\xi )-\frac{1}{2}\ \hat{\xi}_{0}^{2} \\
&=&\int_{0}^{t}\{\langle \hat{\xi},\mathrm{d}\hat{\vartheta}+\frac{1}{2}\ 
\dot{\mathbf{p}}\hat{\xi}\mathrm{d}r\rangle +\langle \mathrm{d}_{-}\hat{\xi},%
\hat{p}(\hat{\xi})\rangle \} \\
&=&\!\int_{0}^{t}\!\{\langle \mathbf{s}\hat{\xi},\mathrm{d}\Upsilon \rangle
-\langle 2\Re \bar{\boldsymbol{\zeta }}\mathrm{d}\hat{\boldsymbol{Y}},\hat{%
\vartheta}\rangle +\!(\langle \boldsymbol{\alpha }\hat{\xi},\mathbf{p}\hat{%
\xi}\rangle +\frac{1}{2}\ \!\dot{p}(\hat{\xi},\hat{\xi})-\bar{\mu}(\mathbf{p}%
\hat{\xi},\mathbf{p}\hat{\xi}))\mathrm{d}r\} \\
&=&\int_{0}^{t}\{\langle \mathbf{s}\hat{\xi},\mathrm{d}\Upsilon \rangle -%
\hat{\vartheta}(2\Re \bar{\boldsymbol{\zeta }})\mathrm{d}\boldsymbol{Y}+%
\frac{1}{2}\ \!\left( \tilde{\varepsilon}(\mathbf{s}\hat{\xi},\mathbf{s}\hat{%
\xi})+\bar{\mu}(\hat{\vartheta},\hat{\vartheta})-\bar{\mu}(\hat{p}(\hat{\xi}%
),\hat{p}(\hat{\xi}))\right) \mathrm{d}r\} \\
&=&\int_{0}^{t}\left\{ \langle \mathbf{s}\hat{\xi},\mathrm{d}\Upsilon
\rangle +\frac{1}{2}\ \left( \tilde{\varepsilon}(\mathbf{s}\hat{\xi},\mathbf{%
s}\hat{\xi})-\bar{\mu}(\hat{p}(\hat{\xi}),\hat{p}(\hat{\xi}))\right) \mathrm{%
d}r\right\} -\ln \hat{\rho}_{t},
\end{eqnarray*}%
which gives~(\ref{eq:ccrintform}) with $\ln \hat{\rho}_{t}=\int_{0}^{t}(\hat{%
\vartheta}(2\Re \bar{\boldsymbol{\zeta }})\mathrm{d}\boldsymbol{Y}-\frac{1}{2%
}\ \bar{\mu}(\hat{\vartheta},\hat{\vartheta})\mathrm{d}r)$.

\textbf{Remark 4.} Let us consider the case of the complex observation (\ref%
{eq:ccr1.10}). Then the stochastic differential equation 
\begin{eqnarray*}
&&i\mathrm{d}\hat{\phi}_{t}(\xi )+\{s(\upsilon _{t},\xi )-\langle 
\boldsymbol{\kappa }_{t}\mathbf{s}\xi ,\partial \rangle +\frac{i}{2}\
\varepsilon _{t}(\mathbf{s}\xi ,\mathbf{s}\xi )\}\hat{\phi}_{t}(\xi )\}%
\mathrm{d}t \\
&=&2\Re \int_{\Lambda }\langle \bar{\zeta}_{t,\mathbf{x}}^{\sharp },\partial 
\hat{\phi}_{t}(\xi )\rangle \mathrm{d}Z_{t}(\mathrm{d}\mathbf{x})+2\Im
\int_{\Lambda }s(\xi ,\bar{\zeta}_{t,\mathbf{x}}^{\sharp })\hat{\phi}%
_{t}(\xi )\mathrm{d}Z_{t}(\mathrm{d}\mathbf{x}),
\end{eqnarray*}%
corresponding to~(\ref{eq:ccr2.10}), has the Gaussian solution~(\ref%
{eq:ccrgausform}) defined by the density 
\begin{equation*}
\hat{\rho}_{t}=\exp \int_{0}^{t}\{2\Re \int_{\Lambda }\hat{\vartheta}_{r}(%
\bar{\zeta}_{r,\mathbf{x}}^{\sharp })\mathrm{d}Z_{r}(\mathrm{d}\mathbf{x})-%
\bar{\varepsilon}_{r}(\hat{\vartheta}_{r},\hat{\vartheta}_{r})\mathrm{d}r\},
\end{equation*}%
where $\bar{\varepsilon}_{t}(\vartheta ,\vartheta )=\int_{\Lambda
}|\vartheta (\bar{\zeta}_{t,\mathbf{x}}^{\sharp })|^{2}\lambda (\mathrm{d}%
\mathbf{x})$, and by the filtering equations 
\begin{equation*}
\mathrm{d}\hat{\vartheta}_{t}(\xi )+\hat{\vartheta}_{t}(i\boldsymbol{\kappa }%
_{t}\mathbf{\ s}\xi )\mathrm{d}t=2\Re \int_{\Lambda }\langle \xi ,\mathbf{k}%
_{t}\bar{\zeta}_{t,\mathbf{x}}^{\sharp }\rangle \mathrm{d}\tilde{Z}_{t}(%
\mathrm{d}\mathbf{x})+\upsilon _{t}(\mathbf{s}\xi )\mathrm{d}t,
\end{equation*}%
where $\mathrm{d}\tilde{Z}_{t}(\mathrm{E})=\mathrm{d}Z_{t}(\mathrm{E})-%
\mathrm{d}t\int_{\mathrm{E}}\hat{\vartheta}_{t}(\bar{\zeta}_{t,\mathbf{x}%
})\lambda (\mathrm{d}\mathbf{x})$. The corresponding complex form of Riccati
equation is 
\begin{equation*}
\frac{\mathrm{d}}{\mathrm{d}t}\ \!p_{t}(\xi ,\xi )+2p_{t}(\xi ,i\boldsymbol{%
\kappa }_{t}\mathbf{s}\xi )=\varepsilon _{t}(\mathbf{s}\xi ,\mathbf{s}\xi )-2%
\bar{\varepsilon}_{t}(\mathbf{k}_{t}\xi ,\mathbf{k}_{t}^{\sharp }\xi ),
\end{equation*}%
where $\mathbf{k}_{t}=\mathbf{p}_{t}+\frac{i}{2}\ \mathbf{s}$, $\mathbf{k}%
_{t}^{\sharp }=\mathbf{p}_{t}-\frac{i}{2}\ \mathbf{s}$.

In the case of the complex Langevin equation (\ref{eq:ccr3.10}),
corresponding to $i\boldsymbol{\kappa }_{t}\mathbf{s}=\boldsymbol{\kappa }%
_{t}\mathbf{c}$ on the invariant subspace $\Xi ^{-}$, the posterior
quasi-free dynamics with complex observation can be described in terms of
the complex parameters $\boldsymbol{\kappa }_{t}=\frac{1}{2}\ \boldsymbol{%
\varepsilon }_{t}+i\boldsymbol{\omega }_{t}$, $\boldsymbol{\kappa }%
_{t}^{\dagger }=\frac{1}{2}\ \boldsymbol{\varepsilon }_{t}-i\boldsymbol{%
\omega }_{t}$, 
\begin{equation*}
\hat{\chi}_{t}=\hat{\vartheta}_{t}|\Xi ^{-},\;\mathbf{l}_{t}=\mathbf{k}%
_{t}^{\sharp }|\Xi ^{-};\;\;\hat{\chi}_{t}(\xi )=0,\;\mathbf{l}_{t}\xi
=0,\;\;\forall \xi \in \Xi ^{+}.
\end{equation*}%
In this case $\hat{\phi}_{t}(\xi \oplus \xi ^{\sharp })=\hat{\rho}_{t}\exp
\{i(\hat{\chi}_{t}^{\sharp }(\xi )+\hat{\chi}_{t}(\xi ^{\sharp }))-\langle
\xi ^{\sharp },\mathbf{p}_{t}\xi \rangle \}$, 
\begin{eqnarray*}
\hat{\rho}_{t} &=&\exp \left\{ \int_{0}^{t}\{2\Re \int_{\Lambda }\hat{\chi}%
_{r}(\bar{\zeta}_{r,\mathbf{x}}^{\sharp })\mathrm{d}Z_{r}(\mathrm{d}\mathbf{x%
})-\bar{\varepsilon}_{r}(\hat{\chi}_{r}^{\sharp },\hat{\chi}_{r})\}\mathrm{d}%
r\right\} , \\
\mathrm{d}\hat{\chi}_{t}(\xi )+\hat{\chi}_{t}(\boldsymbol{\kappa }_{t}%
\mathbf{c}\xi )\mathrm{d}t &=&\int_{\Lambda }\langle \mathbf{l}_{t}\xi ,\bar{%
\zeta}_{t,\mathbf{x}}^{\sharp }\rangle \mathrm{d}\tilde{Z}_{t}(\mathrm{d}%
\mathbf{x})+\eta _{t}(\mathbf{c}\xi )\mathrm{d}t, \\
\frac{\mathrm{d}}{\mathrm{d}t}\ \!\mathbf{p}_{t}+\mathbf{p}_{t}\boldsymbol{%
\kappa }_{t}\mathbf{c}+\mathbf{c}\boldsymbol{\kappa }_{t}^{\dagger }\mathbf{p%
}_{t} &=&\frac{1}{2}\ \mathbf{c}\boldsymbol{\varepsilon }_{t}\mathbf{c}-(%
\mathbf{p}_{t}-\frac{1}{2}\ \mathbf{c})\bar{\boldsymbol{\varepsilon }}_{t}(%
\mathbf{p}_{t}-\frac{1}{2}\ \mathbf{c}),
\end{eqnarray*}%
where $\bar{\boldsymbol{\varepsilon }}_{t}=\bar{\boldsymbol{\kappa }}_{t}+%
\bar{\boldsymbol{\kappa }}_{t}^{\dagger }$ on $\Xi ^{-}$, $\mathbf{p}_{t}=%
\mathbf{l}_{t}+\frac{1}{2}\ \mathbf{c}$ and%
\begin{equation*}
\mathrm{d}\tilde{Z}_{t}(\mathrm{E})=\mathrm{d}Z_{t}(\mathrm{E})-\mathrm{d}%
t\int_{\mathrm{E}}\hat{\chi}_{t}(\bar{\zeta}_{t,\mathbf{x}})\lambda (\mathrm{%
d}\mathbf{x}).
\end{equation*}

\textbf{Example}. Let us consider the stationary case $\boldsymbol{%
\varepsilon }_{t}=\boldsymbol{\varepsilon }$, $\boldsymbol{\omega }_{t}=%
\boldsymbol{\omega }$ and a complete observation when $\mathcal{A=B}$, $\bar{%
\boldsymbol{\varepsilon }}_{t}=\boldsymbol{\varepsilon }$ is invertible and $%
\boldsymbol{\omega }$ satisfies the $\mathbf{c}$-commutativity condition $%
\boldsymbol{\omega }\mathbf{c}\boldsymbol{\varepsilon }=\boldsymbol{%
\varepsilon }\mathbf{c}\boldsymbol{\omega }$ with $\boldsymbol{\varepsilon }$%
. One can easily prove that the last equation describes the continuous
collapse of a posterior state to a Gaussian pure (coherent) state $\phi
_{\infty }$ with the minimal uncertainty $\mathbf{p}_{\infty }=\frac{1}{2}\ 
\boldsymbol{\varepsilon }^{-1}|\boldsymbol{\varepsilon }\mathbf{c}|$, where 
\begin{equation*}
\boldsymbol{\varepsilon }^{-1}|\boldsymbol{\varepsilon }\mathbf{c}|=%
\boldsymbol{\varepsilon }^{-1/2}|\tilde{\mathbf{c}}|\boldsymbol{\varepsilon }%
^{-1/2}=|\mathbf{c}\boldsymbol{\varepsilon }|\boldsymbol{\varepsilon }%
^{-1},\;\tilde{\mathbf{c}}=\boldsymbol{\varepsilon }^{1/2}\mathbf{c}%
\boldsymbol{\varepsilon }^{1/2},\;|\tilde{\mathbf{c}}|=(\tilde{\mathbf{c}}%
^{2})^{1/2}.
\end{equation*}

Indeed, if $\boldsymbol{\omega }\mathbf{c}\boldsymbol{\varepsilon }=%
\boldsymbol{\varepsilon }\mathbf{c}\boldsymbol{\omega }$, then $\boldsymbol{%
\omega }|\mathbf{c}\boldsymbol{\varepsilon }|=|\boldsymbol{\varepsilon }%
\mathbf{c}|\boldsymbol{\omega }$, because from the commutativity of $\tilde{%
\mathbf{c}}$ with $\boldsymbol{\varepsilon }^{-1/2}\boldsymbol{\omega }%
\boldsymbol{\varepsilon }^{-1/2}$ there follows the commutativity of $|%
\tilde{\mathbf{c}}|$ with $\boldsymbol{\varepsilon }^{-1/2}\boldsymbol{%
\omega }\boldsymbol{\varepsilon }^{-1/2}$ and%
\begin{equation*}
\boldsymbol{\omega }|\mathbf{c}\boldsymbol{\varepsilon }|=\boldsymbol{\omega
\varepsilon }^{-1/2}|\boldsymbol{\varepsilon }^{1/2}\mathbf{c}\boldsymbol{%
\varepsilon }^{1/2}|\boldsymbol{\varepsilon }^{1/2}=\boldsymbol{\varepsilon }%
^{1/2}|\boldsymbol{\varepsilon }^{1/2}\mathbf{c}\boldsymbol{\varepsilon }%
^{1/2}|\boldsymbol{\varepsilon }^{-1/2}\boldsymbol{\omega }=|\boldsymbol{%
\varepsilon }\mathbf{c}|\boldsymbol{\omega }.
\end{equation*}%
Hence the Riccati equation 
\begin{equation*}
\frac{\mathrm{d}}{\mathrm{d}t}\ \!\mathbf{p}_{t}+i(\mathbf{p}_{t}\boldsymbol{%
\omega }\mathbf{c}-\mathbf{c}\boldsymbol{\omega }\mathbf{p}_{t})+\mathbf{p}%
_{t}\boldsymbol{\varepsilon }\mathbf{p}_{t}=\frac{1}{4}\ \mathbf{c}%
\boldsymbol{\varepsilon }\mathbf{c},
\end{equation*}%
corresponding to $\tilde{\varepsilon}=\boldsymbol{\varepsilon }-\bar{%
\boldsymbol{\varepsilon }}=0$, has the unique stationary positive solution $%
\mathbf{p}_{\infty }:\mathbf{p}_{\infty }\boldsymbol{\omega }\mathbf{c}=%
\mathbf{c}\boldsymbol{\omega }\mathbf{p}_{\infty },\;\mathbf{p}_{\infty }%
\boldsymbol{\varepsilon }\mathbf{p}_{\infty }=\frac{1}{4}\ \!\boldsymbol{%
\varepsilon }^{-1/2}|\tilde{\mathbf{c}}|^{2}\boldsymbol{\varepsilon }^{-1/2}=%
\frac{1}{4}\ \mathbf{c}\boldsymbol{\varepsilon }\mathbf{c}$. The convergence 
$\mathbf{p}_{t}\rightarrow \mathbf{p}_{\infty }$ follows from properties of
the Riccati equation with unique stationary solution $\mathbf{p}_{\infty }>0$%
. Thus in the case $\boldsymbol{\varepsilon }=\varepsilon \mathbf{1}$ and $%
\boldsymbol{\omega }=\omega \mathbf{1}$ the positive solution $\mathbf{p}%
_{t} $ corresponding to $\mathbf{p}_{0}=\mathbf{1}$ has the form 
\begin{equation*}
\mathbf{p}_{t}=\frac{|\mathbf{c}|}{2}\ \frac{1+\mathbf{q}_{t}}{1-\mathbf{q}%
_{t}}\ ,\;\mathbf{q}_{t}=\mathbf{q}_{0}\exp \{-\varepsilon |\mathbf{c}|t\},\;%
\mathbf{q}_{0}=\frac{2-|\mathbf{c}|}{2+|\mathbf{c}|}\ ,
\end{equation*}%
and $\mathbf{p}_{\infty }=\frac{1}{2}\ |\mathbf{c}|$, $\mathbf{p}_{t}\approx 
\mathbf{p}_{\infty }+|\mathbf{c}|\mathbf{q}_{0}e^{-\varepsilon |\mathbf{c}%
|t} $ for $t\gg \frac{1}{\varepsilon }\ $, if $|\mathbf{c}|>0$. Hence $%
\mathbf{p}_{t}=\mathbf{p}_{\infty }$ only in the purely quantum case $|%
\mathbf{c}|=2$, and $\mathbf{p}_{t}\rightarrow 0$ only in the purely
classical case $\mathbf{c}=0$ when $\mathbf{p}_{t}=\mathbf{1}/(1+\varepsilon
t)$.

This result was obtained in \cite{bib:b1} for the case of positive definite $%
\mathbf{c}>0$ (a stable quantum system), when the stationary quantum linear
filter 
\begin{equation*}
\mathrm{d}\hat{\chi}_{t}(\xi )+\hat{\chi}_{t}(\kappa \mathbf{c}\xi )\mathrm{d%
}t=\int_{\Lambda }\langle \mathbf{l}\xi ,\zeta _{x}^{\sharp }\rangle (%
\mathrm{d}Z_{t}(\mathrm{d}\mathbf{x})-\hat{\chi}_{t}(\zeta _{x})\lambda (%
\mathrm{d}\mathbf{x})\mathrm{d}t),
\end{equation*}%
corresponding to $\zeta _{t,\mathbf{x}}=\zeta _{x}$, $\kappa =\frac{1}{2}\
\varepsilon +i\omega $, $\mathbf{l}=\frac{1}{2}\ (|\mathbf{c}|-\mathbf{c})$
and $\upsilon _{t}=0$, does not depend on the observable process $Z_{t}:%
\mathbf{l}=0$, if $\mathbf{c}\geq 0$. The a posterior state for the Gaussian
initial wave function $\psi _{0}$ tends asymptotically to the ground state
even without observation as the coherent a priori state: $\hat{\chi}_{t}(\xi
)=\chi _{0}(e^{-\kappa \mathbf{c}t}\xi )\rightarrow 0$.

In the contrary case $\mathbf{c}<0$, $\mathbf{l}=|\mathbf{c}|$,
corresponding to the unstable system, the complete nondemolition observation
is needed to keep the system in a state with the minimal uncertainty
relation. This explains why the quantum open (unstable) oscillator,
corresponding \cite{bib:ref2} to the case $\Xi ^{-}=\mathbb{C}=\Xi ^{+}$
with $|\mathbf{c}|=2$, tends asymptotically to the pure Gaussian state under
the continuous observation of its amplitude $L=R(\zeta )$, given by the
measurement of the complex nondemolition process $Z(t)=L(t)+W(t)$.

\textbf{Acknowledgements.} I am grateful for the hospitality and stimulating
discussions to Professor R. Hudson from Nottingham University where this
paper was begun and to Professor W. von Waldenfels from Heidelberg
University where it was finished.

\section{Conclusion}

The developed time-continuous quantum measurement theory based on the
nondemolition principle as a superselection rule for the output observable
processes abandons the von Neumann projection postulate as a redundancy in
the stochastic extension of the quantum theory. It treats the wave packet
reduction not as a real dynamical process but rather as the statistical
inference of the posterior state described by the conditional wave-function
for the prediction of the probabilities of the future measurements
conditioned by the past observations.

There is no need to postulate in quantum stochastic theory a nonunitary
stochastic linear or nonlinear evolution for the continuous state-vector
reduction as it is done in the phenomenological quantum theories of quantum
trajectories, state diffusion or spontaneous localization \cite{bib:bel6,
bib:bel7, bib:bel8, bib:bel9, bib:bel10}. The nonunitary classical
stochastic evolution giving the continuous reduction and localization of the
posterior state can be rigorously derived \cite{bib:2, bib:b7, bib:bel2,
bib:bel3} within the quantum stochastic unitary evolution \ of the
correspondent compound system, the object of the measurement and the input
Bose field in the vacuum state as it is shown here.


\end{document}